\documentclass{article}
\usepackage{graphicx}
\usepackage{epsfig}
\usepackage{amsfonts}
\usepackage{cite}
 \usepackage{amssymb}

\date{\today}

\newcommand{\insertplot}[5]{\begin{figure}
 \hfill\hbox to 0.05in{\vbox to #5in{\vfill
 \inputplot{#1}{#4}{#5}}\hfill}
 \hfill\vspace{-.1in}
 \caption{#2}\label{#3}
 \end{figure}}
 \newcommand{\inputplot}[3]{
 \special{ps: plotfile #1}
}
\newcounter{fig}

\newcommand{\ee}{\end{equation}}
\newcommand{\eea}{\end{eqnarray}}
\newcommand{\be}{\begin{equation}}
\newcommand{\bea}{\begin{eqnarray}}

\newcommand{\pa}{\partial}

\begin{document}

\title{ {\bf Dynamically and thermodynamically stable black holes \\ in Einstein-Maxwell-dilaton gravity}}
 \vspace{1.5truecm}
\author{
{\bf Dumitru Astefanesei$^1$},
{\bf Jose Luis Bl\'azquez-Salcedo$^2$},
{\bf Carlos Herdeiro$^3$}, \\
{\bf Eugen Radu$^{3}$}
and
{\bf Nicolas Sanchis-Gual$^4$}
\\
\\
%
{\small $^1$ Pontificia Universidad Cat\'olica de
	Valpara\'\i so, Instituto de F\'{\i}sica
	},
 \\ 
{\small Av. Brasil 2950, Valpara\'{\i}so, Chile}
\\
{\small $^2$ Institut f\"ur  Physik, Universit\"at Oldenburg, Postfach 2503,
D-26111 Oldenburg, Germany}
\\
{\small 
$^3$
Departamento de Matem\'atica  da Universidade de Aveiro and}
\\
{\small 
 Center for Research and Development in Mathematics and Applications (CIDMA),
}
\\
{\small 
 Campus de Santiago, 3810-183 Aveiro, Portugal
}
\\
{\small 
$^4$
Centro de Astrof\'\i sica e Gravitac\~ao - CENTRA, Departamento de F\'\i sica,
Instituto Superior T\'ecnico - IST,}
\\
{\small Universidade de Lisboa - UL, Avenida
Rovisco Pais 1, 1049-001, Portugal} 
}

\date{December 2019}

\maketitle

\begin{abstract}
We consider Einstein-Maxwell-dilaton gravity with the non-minimal exponential coupling between the dilaton and the Maxwell field emerging from low energy heterotic string theory. The dilaton is endowed with a potential that originates from an electromagnetic Fayet-Iliopoulos (FI) term in $\mathcal{N}=2$ extended supergravity in four spacetime dimensions.  For the case we are interested in, this potential introduces a single parameter $\alpha$. When $\alpha\rightarrow 0$, the static black holes  (BHs) of the model are the Gibbons-Maeda-Garfinkle-Horowitz-Strominger (GMGHS) solutions. When $\alpha \rightarrow \infty$, the BHs become the standard Reissner-Nordstr\"om (RN) solutions of electrovacuum General Relativity. The BH solutions for finite non-zero $\alpha$ interpolate between these two families. In this case, the dilaton potential regularizes the extremal limit of the GMGHS solution yielding a set of zero temperature BHs with a near horizon $AdS_2\times S^2$ geometry. We show that, in the neighborhood of these extremal solutions, there is a subset of BHs that are dynamically and thermodynamically stable, all of which have charge to mass ratio larger than unity. By dynamical stability we mean that no growing quasi-normal modes are found; thus they are stable against linear perturbations (spherical and non-spherical). Moreover, non-linear numerical evolutions lend support to their non-linear stability. By thermodynamical stability we mean the BHs are stable both in the canonical and grand-canonical ensemble. In particular, both the specific heat at constant charge and the isothermal permittivity are positive. This is not possible for RN and GMGHS BHs. We discuss the different thermodynamical phases for the BHs in this model and comment on what may allow the existence of both dynamically and thermodynamically stable BHs.
\end{abstract}

\newpage

\tableofcontents

\section{Introduction}
Stable configurations are the preferred configurations in (thermo)dynamics. For BHs, the classical concept of stability is that of dynamical (or mechanical) stability. Is the BH robust against small, mechanical perturbations? Perturbative stability of BHs is typically assessed by computing the spectrum of quasi-normal modes (QNMs) of the isolated BH. The absence of growing modes establishes mode stability.  For the only static vacuum BH of General Relativity, the Schwarzschild solution~\cite{Israel:1967wq}, its mode stability was established by the seminal works of Regge and Wheeler~\cite{Regge:1957td} and Zerilli~\cite{Zerilli:1970se}. In this sense, the Schwarzschild BH is a preferred configuration. 

The advent of BH thermodynamics in the 1970s~\cite{Hawking:1974sw,Hawking:1976de}, as a consequence of a semi-classical treatment of gravity, yields a different angle on BH stability. Is a BH in thermodynamical equilibrium with its environment robust against small fluctuations of, say, energy, or another of its defining parameters?  This question is, a priori, different, because the BH is interacting with an environment, or reservoir, rather than being isolated. Different reservoirs, or statistical ensembles, can be considered. For the case of  a Schwarzschild BH, whose only defining parameter is its mass, the question simplifies. Fixing the reservoir's temperature, one simply asks how does the BH temperature responds to a small fluctuation of the BH energy. It turns out that the BH heats up/cools down when it loses/absorbs energy. That is, it has a negative specific heat. Thus, under a small energy exchange, Schwarzschild BHs run away from thermal equilibrium when placed in a reservoir at fixed temperature. They are (locally) thermodynamically unstable. Consequently, in this sense, these  BHs are not preferred configurations. 

Consider now the addition of electric charge. In electrovacuum General Relativity, the only static electrically charged BH solution with a connected horizon is the Reissner-Nordstr\"om (RN) BH~\cite{Israel:1967za}. Dynamically, it is still perturbatively stable, since its QNMs decay in time~\cite{Moncrief:1974gw,Moncrief:1974ng}. Thermodynamically, however, one can now consider different interactions. Firstly, consider the BH can exchange energy, but not electric charge, with the reservoir. Thus, the BH charge is fixed. Is the heat capacity (at constant charge) still negative? It must be for small charge, as the solution reduces to the Schwarzschild BH. For sufficiently large charge to mass ratio, $q\equiv Q_e/M> \sqrt{3}/2$, however, the specific heat (at constant charge) becomes positive~\cite{Davies:1977}. Thus, preventing any charge exchanges,  RN BHs with sufficiently large $q$ are (locally) thermodynamically stable, oscillating around the reservoir's temperature when small exchanges of energy occur. This is the canonical ensemble. Under these conditions, RN BHs with $q> \sqrt{3}/2$ are preferred. 

Considering no electric charge exchanges is, however, non-generic. In a generic situations such exchanges do occur. The reservoir is now not only a reservoir of energy, kept at constant temperature, but also of (charged) particles, kept at constant ``chemical"  potential, which in this case corresponds to the electrostatic potential. This is the grand-canonical ensemble. Now, the assessment of stability must also consider a possible run away mode triggered by the wrong evolution of the BH's chemical (electrostatic) potential. As the BH absorbs (releases) positive charge, if its chemical potential decreases (increases), this promotes more charge  absorption (release), and hence a run-away mode. The response function monitoring this effect is the isothermal permittivity (at constant temperature). It so happens that for RN BHs it becomes negative precisely for $q>\sqrt{3}/2$, exactly when the heat capacity (at constant charge) becomes positive. This can be understood as follows. Fixing the temperature, an increase of charge implies an (even larger) increase of the BH mass, for $q>\sqrt{3}/2$. This implies the BH size increases sufficiently so that, despite the charge increase, the electrostatic potential decreases. The bottom line is that RN BHs are always locally unstable in the grand canonical ensemble. For small charges, exchange of energy at constant charge (or at constant electrostatic potential) promotes a run-away mode. For large charges, exchange of charge at constant temperature promotes the instability. Thus, under these conditions, RN BHs are not preferred configurations. Intriguingly, however, the isothermal permittivity diverges as extremality ($q\rightarrow 1$) is approached, suggesting the RN family is on the verge of another thermodynamical phase.

The advent of supergravity and string theory naturally led to considering Einstein-Maxwell models with an extra scalar field, a dilaton, non-minimally coupled to the Maxwell field with a particular exponential coupling. In fact, such models naturally occur also in the context of Kaluza-Klein theories. Gibbons~\cite{Gibbons:1982ih}, subsequently also with Maeda~\cite{Gibbons:1987ps}, considered the charged BH solutions in these models. They are charged BHs that possess scalar (dilaton) ``hair". The BH solutions of these Einstein-Maxwell-dilaton models were later reobtained by Garfinkle, Horowitz and Strominger in the context of string theory~\cite{Garfinkle:1990qj}. We shall refer to them as GMGHS solutions.  

The GMGHS BHs, with the particular coupling emerging from string theory, are sometimes regarded as a sort of generalisation of the RN BH of electrovacuum, albeit  they do not reduce to the latter, except in the uncharged limit. Moreover, GMGHS BHs introduce three qualitative new physical aspects. Firstly, the excited dilaton around the BH creates an effective medium with electric properties, which can be faced as, say, inducing a varying magnetic permeability.
Secondly, in the standard electrovacuum-like description (the Einstein frame), the GMGHS BHs do not have a smooth extremal limit; they are singular in that limit. Thirdly, these BHs allow a charge to mass ratio greater than unity. 

The GMGHS BHs can be made more RN-like, in particular concerning the second property of the previous paragraph, by augmenting the model with a particular dilaton potential that has been shown to  emerge in $\mathcal{N}=2$ supergravity in four spacetime dimensions, extended with vector multiplets and deformed by a Fayet-Iliopoulos term~\cite{Anabalon:2017yhv}. In~\cite{Anabalon:2013qua}, Anabalon, Astefanesei and Mann obtained exact charged BH solutions in this model, that we shall refer to as AAM BHs. The potential introduces one single extra parameter, $\alpha$. Then, as $\alpha\rightarrow 0$, the stringy GMGHS BHs are recovered. On the other hand, as $\alpha\rightarrow \infty$, the potential confines the scalar field to vanish but the electric charge may remain non-zero, yielding the RN electrovacuum family. In this sense, the AAM family of BHs interpolates between the RN and the GMGHS families. Now, it turns out that  the dilatonic potential regularises the extremal limit, yielding a family of extremal BHs with a near horizon geometry of Robinson-Bertotti type, $AdS_2\times S^2$, analogue to the extremal RN solution.
Nonetheless, the AAM solution retains the GMGHS property that it still allows overcharged BHs with $q=1$. The AAM family of solutions, therefore, presents itself as an arena to test the hypothesis that the $q\rightarrow 1$ RN BHs are on the verge of another thermodynamical phase. 

In this work we therefore investigate the dynamical and thermodynamical stability of the AAM BHs. It was recently pointed out~\cite{Astefanesei:2019mds} that some of these BHs are thermodynamically stable in both the canonical and grand-canonical ensemble, unlike RN BHs. In this paper we perform a thorough scanning of the domain of existence, precisely identifying the subset of thermodynamically stable AAM BHs. Our analysis makes clear that this only occurs in the overcharged regime, $q>1$. 
Indeed, the AAM solutions have three thermodynamical phases: 
{\bf i)} a Schwarzschild-like phase, which is unstable since the specific heat (at constant charge or at constant electrostatic potential) is negative; 
{\bf ii)} a near-extremal RN-like phase, which is unstable since the isothermal permittivity is negative; 
and 
{\bf iii)} a new stable phase, for which a necessary, but not sufficient,  condition is that the BH is overcharged. In this sense the AAM BHs are an extension of RN BHs into the overcharged regime, which is made possible by the conjugation of the dilaton non-minimal coupling and potential. 

Our work also clarifies that, like RN BHs, GMGHS BHs are never thermodynamically stable in both the canonical and grand-canonical ensemble. Moreover, we show that, like RN BHs, the AAM family is dynamically perturbatively stable, in the sense that no growing quasi-normal modes are found, building upon the recent work~\cite{Blazquez-Salcedo:2019nwd}, see also~\cite{Jansen:2019wag}. The dynamical robustness of AAM solutions is furthermore confirmed by fully non-linear numerical simulations with the Einstein-Maxwell-dilaton system. Even starting with initial data far off from the equilibrium solutions, the evolutions relax to the latter. These simulations were performed using a similar code and setup to the ones reported in~\cite{Herdeiro:2018wub,Fernandes:2019rez,Fernandes:2019kmh}. All this put together means that there is a subset of AAM solutions that are preferred, both dynamically and thermodynamically. To the best of our knowledge, this is the first such example for asymptotically flat BHs, without using artefacts such as box boundary conditions. 

This paper is organised as follows. In Section~\ref{sec2} we describe the model and the AAM family of solutions. In particular we consider the extremal limit showing the existence of $AdS_2\times S^2$ geometries, unlike in the GMGHS sub-family. In Section~\ref{sec3} we present a linear analysis of dynamical stability. We consider separately spherical, axial and polar perturbations. Then we discuss the spectrum of QNMs showing that, within our scanning, they always decay in time, providing strong evidence of mode stability. In Section~\ref{sec4} we consider a non-linear analysis of dynamical stability, showing that fully non-linear dynamical evolutions within the Einstein-Maxwell-dilaton model, starting from initial data which can be considered a highly perturbed AAM BH, the evolution converges to the latter. In Section~\ref{sec5} we discuss thermodynamical stability in both the canonical and grand-canonical ensemble, spelling out the conditions and identifying the region of the domain of existence of AAM BHs where thermodynamical stability holds. Finally, in Section~\ref{sec6} we summarise our results and provide a discussion of their significance.

\section{Einstein-Maxwell-dilaton BHs with a supergravity potential}
\label{sec2}
%
\subsection{The model}
\begin{figure}[t]
\centering
\includegraphics[width=0.55\linewidth,angle=0]{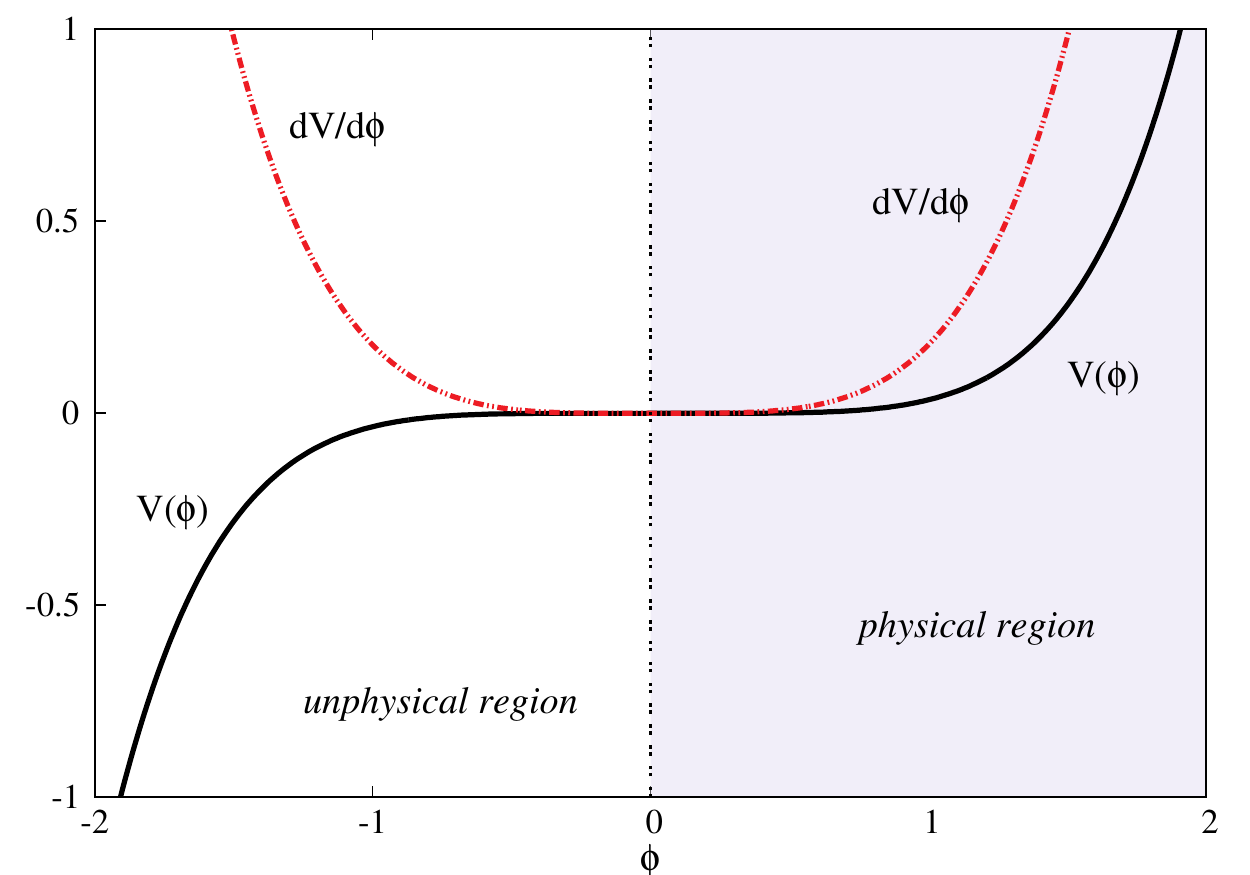}
\caption{The potential of the model, eq.~(2.2).
}
\label{Fig0}
\end{figure}

The model under consideration is an Einstein-Maxwell-dilaton model endowed with a particular dilaton potential. It is described by the action:
\begin{eqnarray}
\label{action}
\mathcal{S} \left[g_{\mu\nu},A_\mu,\phi\right]
=\frac{1}{16 \pi G}\int 
{d^{4}x\sqrt{-g}}\left[
R
-e^{\gamma\phi} F_{\mu\nu}F^{\mu\nu}
-\frac{1}{2}\pa_\mu\phi\,\pa^\mu\phi
-V(\phi)\right] \ .
\end{eqnarray} 
Here, $R$ is the Ricci scalar,  $F_{\mu \nu}\equiv\partial_\mu A_\nu-\partial_\nu A_\mu$ is the Maxwell field and  $\phi$ is the dilaton field, which couples non-minimally to the Maxwell field with a coupling constant $\gamma$. In the following, we use units where $8\pi G=1=c$. The dilaton is endowed with the potential discussed in~\cite{Astefanesei:2019mds},
which has the form:
\begin{eqnarray}
\label{V}
V(\phi) = 2\alpha(2\phi+\phi \cosh\phi-3\sinh\phi) \ ,
\end{eqnarray}
where $\alpha$
is a dimensionful constant, with dimensions ${\rm length}^{-2}$. This potential is plotted in Fig.~\ref{Fig0}. For small $\phi$, this potential behaves $V \simeq  \alpha \phi^5/30$; thus it contains no mass term. 
 We notice that $V$ is invariant under the discrete symmetry
$\alpha\to -\alpha $,
$\phi \to -\phi$.
Thus,
taking $\alpha>0$ without any loss of generality, 
the  requirement $V>0$
imposes that the physical 
solutions necessarily have
a positive $\phi$.

%
The equations of motion that follow from~(\ref{action}) are 
\begin{eqnarray}
\label{eins} 
&&
R_{\mu\nu}-\frac{1}{2}g_{\mu\nu}R=
\frac{1}{2}
\left(
T_{\mu\nu}^{(\phi)}+T_{\mu\nu}^{(M)}
\right) \ , \qquad 
\partial_{\mu}(\sqrt{-g}e^{\gamma\phi}F^{\mu\nu} )
=0 \ ,
\\
&&
\frac{1}{\sqrt{-g}}\partial_{\mu}
 (\sqrt{-g}g^{\mu\nu}\partial_{\nu}\phi )
 =\frac{dV(\phi)}{d\phi}+\gamma{} {e}^{\gamma\phi}F^2 \ ,
\label{klein}
\end{eqnarray}
where 
\begin{eqnarray}
T_{\mu\nu}^{(\phi)}
\equiv \partial_{\mu}\phi\partial_{\nu}\phi
- g_{\mu\nu}
 \left[ \frac{1}{2}(\partial \phi)^2+V(\phi)  \right] \ , \qquad 
T_{\mu\nu}^{(M)}\equiv 4e^{\gamma\phi}
\left(F_{\mu\alpha}F_{\nu}^{\,\,\alpha}
-\frac{1}{4}g_{\mu\nu}F^2\right) \ ,
\end{eqnarray}
are the dilaton and electromagnetic energy-momentum tensors. 

The potential (\ref{V}) was first considered in~\cite{Anabalon:2013qua}. It was originally engineered to obtain exact solutions in Einstein-Maxwell-dilaton gravity. Subsequently, it was shown that the Einstein-dilaton sector of~(\ref{action}) is a consistent truncation of $\mathcal{N}=2$ supergravity in four spacetime dimensions, coupled to a vector multiplet and deformed by a Fayet-Iliopoulos term~\cite{Anabalon:2017yhv}. Augmenting the latter model by introducing the standard Maxwell term with the dilatonic coupling that emerges in low energy heterotic string theory~\cite{Garfinkle:1990qj} leads to the action~(\ref{action}). It is possible, but it has not been explicitly shown, that the full model~(\ref{action}) emerges from supergravity. That is an interesting open question. But both its $F=0$ and $\alpha=0$ truncations emerge from supergravity models.

\subsection{The solutions in Schwarzschild-like coordinates}
The charged, spherically symmetric BH solutions of the model~(\ref{action}) are the standard RN BH for $ \gamma=0=\alpha$, 
the GMGHS solution~\cite{Garfinkle:1990qj} for $\alpha=0$ and arbitrary $\gamma$, and the AAM BHs~\cite{Anabalon:2013qua} 
for  $\gamma=1$ and arbitrary $\alpha$. The last case will be the focus of our work. Obviously, the AAM solutions reduce to the subset of the GMGHS solutions with $\gamma=1$ when $\alpha=0$. But, as shown below, they turn out  also to reduce to RN BHs when $\alpha\rightarrow \infty$. 

As discussed in \cite{Astefanesei:2019mds},
these solutions form two branches 
distinguished by the sign of the parameter $\alpha$ in
the dilaton potential (\ref{V}).
In what follows we shall take $\alpha>0$. Then, following~\cite{Astefanesei:2019mds}, there is a family of BH solutions with strictly positive $\phi$, thus only probing the strictly positive region of the dilaton potential. 
%
Employing Schwarzschild-like coordinates\footnote{In~\cite{Astefanesei:2019mds} the solution was considered in a different coordinate system, which makes its properties less transparent. 
The relation between the radial  $x-$coordinate in
\cite{Astefanesei:2019mds}
and the $r$-coordinate herein is
$x=(1+2\eta^2 r^2+\sqrt{1+4 \eta^2 r^2})/(2\eta^2 r^2)$,
with  $\eta=1/Q_s$. 
Also, the
parameter $q$ in Ref. \cite{Astefanesei:2019mds}
should not be confused with $q=Q_e/M$ $cf.$ (\ref{scale1}) in this work.  
}, the geometry of these BH solutions takes the form
\begin{eqnarray}
\label{s1}
ds^2=-N(r)\sigma^2(r) dt^2+\frac{dr^2}{N(r)}+r^2 (d\theta^2+\sin^2\theta d\varphi^2)\ , ~~~
{\rm where}~~N(r)\equiv 1-\frac{2m(r)}{r} \ ,
\end{eqnarray}
where  the Misner-Sharp  mass function $m(r)$~\cite{Misner:1964je}
and  the redshift function $\sigma(r) $ read 
\begin{equation}
\label{s2}
m(r)=\frac{r}{2\sigma^2(r)}
\left\{\frac{Q_e^2}{r^2}[\chi(r)-1]
-\alpha 
\left[
\frac{Q_s^2}{2}\chi(r)-r^2 \phi(r)
\right] \right\}-\frac{Q_s^2}{8r}\ , ~~~
\sigma(r)=\left(1+\frac{Q_s^2}{4r^2}\right)^{-1/2} \ ,
\end{equation}
where
\begin{equation}
\chi(r)\equiv \sqrt{1+\frac{4r^2}{Q_s^2}} \ ,
\end{equation}
while the  matter fields take the form:
\begin{eqnarray}
\label{s3}
e^{\phi(r)/2}= \frac{Q_s}{2r}[1+\chi(r)] \ , ~~~ {\rm and}~~~
\label{s4}
A=a_0(r) dt \ , ~~~  {\rm where}~~~
a_0(r)=\frac{Q_e Q_s}{2r^2}[1-\chi(r)]+A_0~; \ \ \ 
\end{eqnarray}
$A_0$ is an arbitrary constant.
Besides $A_0$, this solution contains two arbitrary parameters 
$Q_e$ and $Q_s$,
which correspond to the 
electric charge and scalar charge respectively,
as read $e.g.$
from the $1/r$ terms in the asymptotic expansion of the scalar and gauge field. 
Observe, however, that $Q_e$ is a gauge charge associated to 
a gauge symmetry, whereas $Q_s$ is not.\footnote{We also emphasise that unlike \cite{Astefanesei:2018vga}, due to the existence of the potential, to obtain asymptotically flat solutions the asymptotic value of the scalar field should be kept fixed.}

The ADM mass is read off from the asymptotic value of $m(r)$; it can be expressed as a function of $Q_e$, $Q_s$ 
\begin{eqnarray}
\label{s5}
 M= \frac{Q_e^2}{Q_s}-\frac{\alpha}{12}Q_s^3 \ .
\end{eqnarray}
Thus, the scalar charge is not independent from the mass and gauge charge. 
We conclude that the scalar hair of the solution is of secondary type --- see~\cite{Herdeiro:2015waa} for a discussion of scalar hair for asymptotically flat BHs.

For a certain parameter range the above solution describes BHs. Then, there is a BH  horizon at $r=r_H>0$. The relation between the electric  charge,\footnote{The electric charge can be computed by the covariant form of Gauss' law
\begin{eqnarray}
\label{rs4s}
Q_e=\frac{1}{8\pi } \oint_S dS_{\mu\nu}e^{\gamma \phi }F^{\mu\nu}~.
\end{eqnarray} 
Since there are no sources to Maxwell's equations~(\ref{eins}) outside the horizon, the value of the charge is independent 
of the choice of the surface $S$, with
$A_t\sim A_0-Q_e/r$ asymptotically. 
The scalar charge $Q_s$ is computed from the asymptotics of the scalar field, $\phi\sim Q_s/r$.
}
scalar charge and $r_H$ reads:
\begin{eqnarray}
\label{rs4}
Q_e=\frac{Q_s}{2 }\sqrt{ 
\left[
1+\chi(r_H)
\right]
\left\{
    1+\frac{\alpha Q_s^2}{2} 
\left[
\chi(r_H)-\frac{2 r_H^2\phi(r_H)}{Q_s^2}
\right]
\right\}
} \ .
\end{eqnarray}
For these BHs,
the Hawking temperature 
$T_H$ 
and
 the
event horizon area 
$A_H$
 read
\begin{eqnarray}
\label{s9} 
&&
\nonumber
T_H= \frac{1}{4\pi}N'(r_H)\sigma(r_H)
=
\frac{Q_s}{8\pi r_H^2}
\left[
\left(1-\frac{\alpha Q_s^2}{2}\right)
[\chi(r_H)-1]
+\alpha r_H^2
\left\{
6+[1-3\chi(r_H)]\phi(r_H)  
\right\}
\right] \ , 
\\
&&
A_H=4\pi r_H^2 \ .
\end{eqnarray}
Also, of relevance for the analysis below, working in a gauge with $A_t(r_H)=0$
we find the electrostatic potential at infinity, which corresponds to the chemical potential  in the thermodynamical analysis, is
\begin{eqnarray}
\label{s8}
\Phi=\frac{2}{1+\chi(r_H)} \frac{Q_e}{Q_s} \ .
\end{eqnarray}
For any choice of the dilaton potential (and a vanishing dilaton at infinity),
one can verify that  
the solutions satisfy the first law of BH thermodynamics in the form
\begin{equation}
\label{1st}
dM=\frac{1}{4}T_H dA_H+\Phi dQ_e \ .
\end{equation}
Also,
one can prove that solutions satisfy the following Smarr-law
\begin{equation}
\label{Smarr}
 M=2T_H S+\Phi Q_e+M_{\phi}\ , \qquad {\rm with}~~M_{\phi}=\int d^3x_{~}V(\phi) \ ,
\end{equation}
where the space integral is taken over a domain bounded by the event horizon
and the sphere at infinity\footnote{The expression (\ref{Smarr})
holds for a generic model (\ref{action}).
For the AAM solution, the explicit form of $M_{\phi}$ reads
\begin{equation}
\label{Mphi}
 M_{\phi}= \alpha Q_s r_H^2\left[\frac{1}{2}\chi(r_H)\phi(r_H)-\frac{Q_s^2}{12r_H^2}-1 \right]\ .
\end{equation}
}.
Moreover, as usual in Einstein gravity, the entropy is given by the Bekenstein-Hawking formula
\begin{equation}
\label{S}
S=\frac{1}{4}A_H \ .
\end{equation}

The model possesses the scaling symmetry
 \begin{eqnarray}
\label{scale} 
 r \to \lambda r \ , \qquad \alpha \to  \alpha/\lambda^2 ,
\end{eqnarray}
where $\lambda>0$ is a constant. 
Under this scaling symmetry, all other quantities scale appropriately, $e.g.$
$M \to \lambda M$, $Q_e \to \lambda Q_e$ and $Q_s \to \lambda Q_s$.
We frame the physical discussion using quantities which are invariant under this transformation.
These are  the following  reduced quantities
\begin{eqnarray}
\label{scale1}
q\equiv \frac{Q_e}{M}\ , \qquad a_H\equiv \frac{A_H}{16\pi M^2}\ , \qquad t_H\equiv 8\pi T_H M \  .
\end{eqnarray}

\begin{figure}[t]
\centering
\includegraphics[width=0.49\linewidth,angle=0]{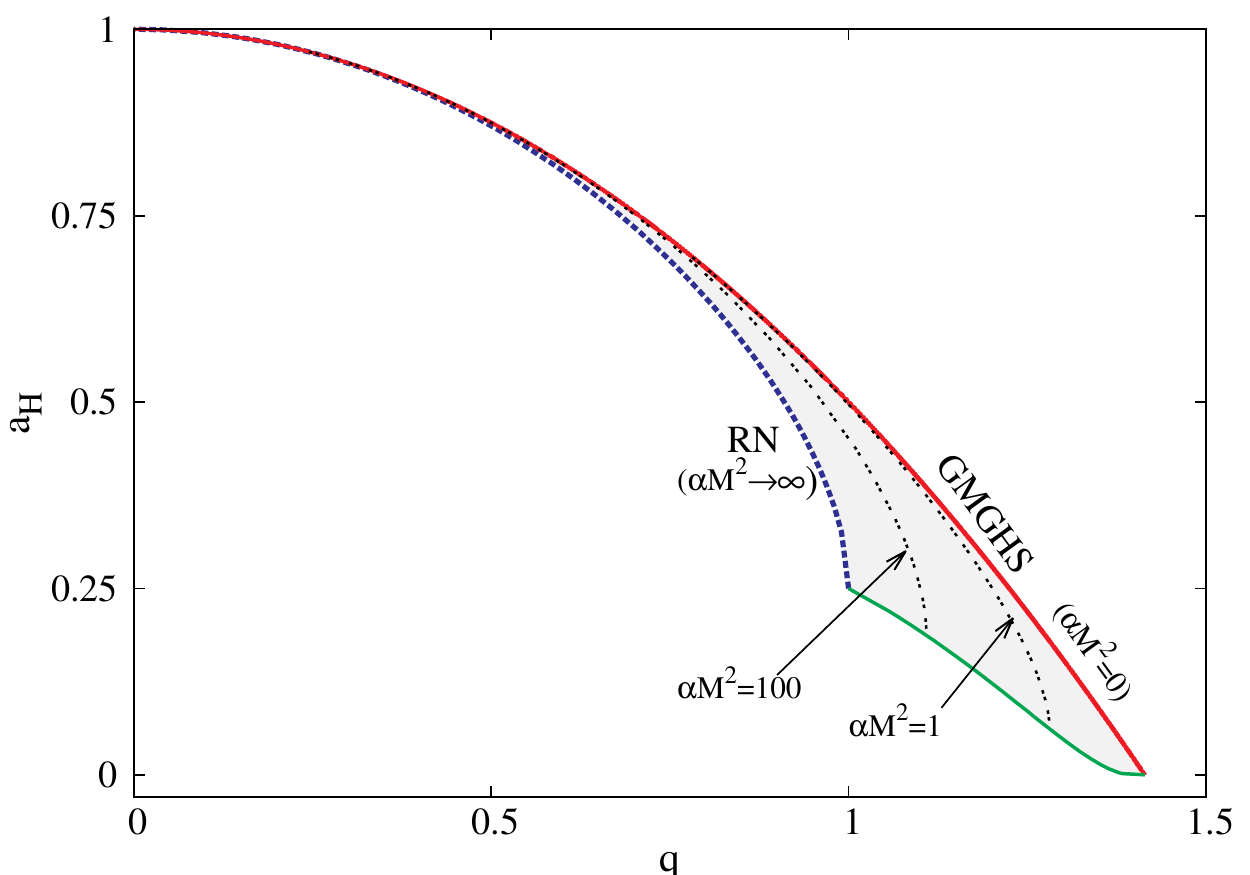}
\includegraphics[width=0.49\linewidth,angle=0]{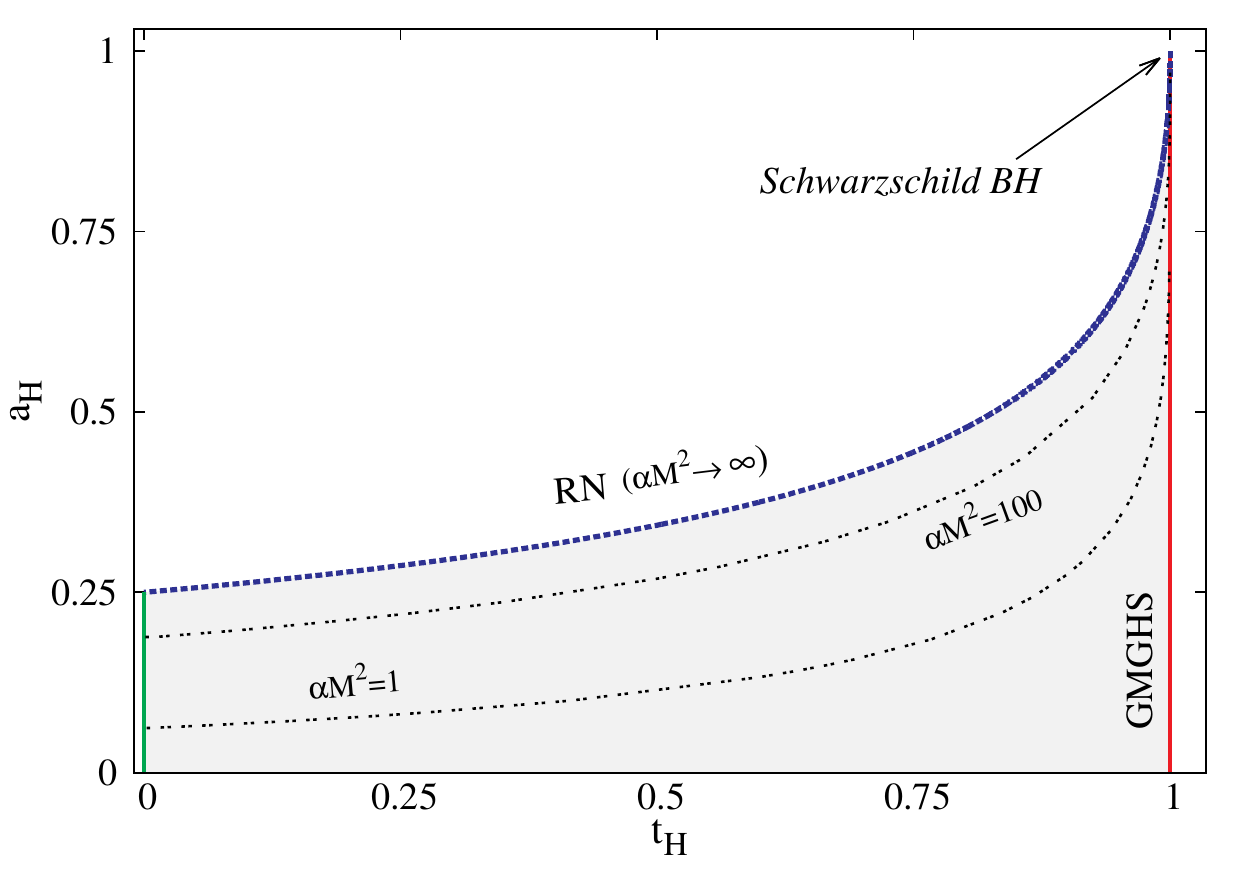}
\caption{Domain of existence of the AAM BH solutions  
in a reduced area $vs.$ $q$  (left panel) or $t_H$ (right panel) diagram - shaded region. The RN (GMGHS) limit is given by the dashed blue (solid red)
lines. Black dashed lines correspond to $\alpha M^2=$constant. The solid green line corresponds to the extremal limit.}
\label{Fig1}
\end{figure}

\begin{figure}[h!]
\centering
\includegraphics[width=0.49\linewidth,angle=0]{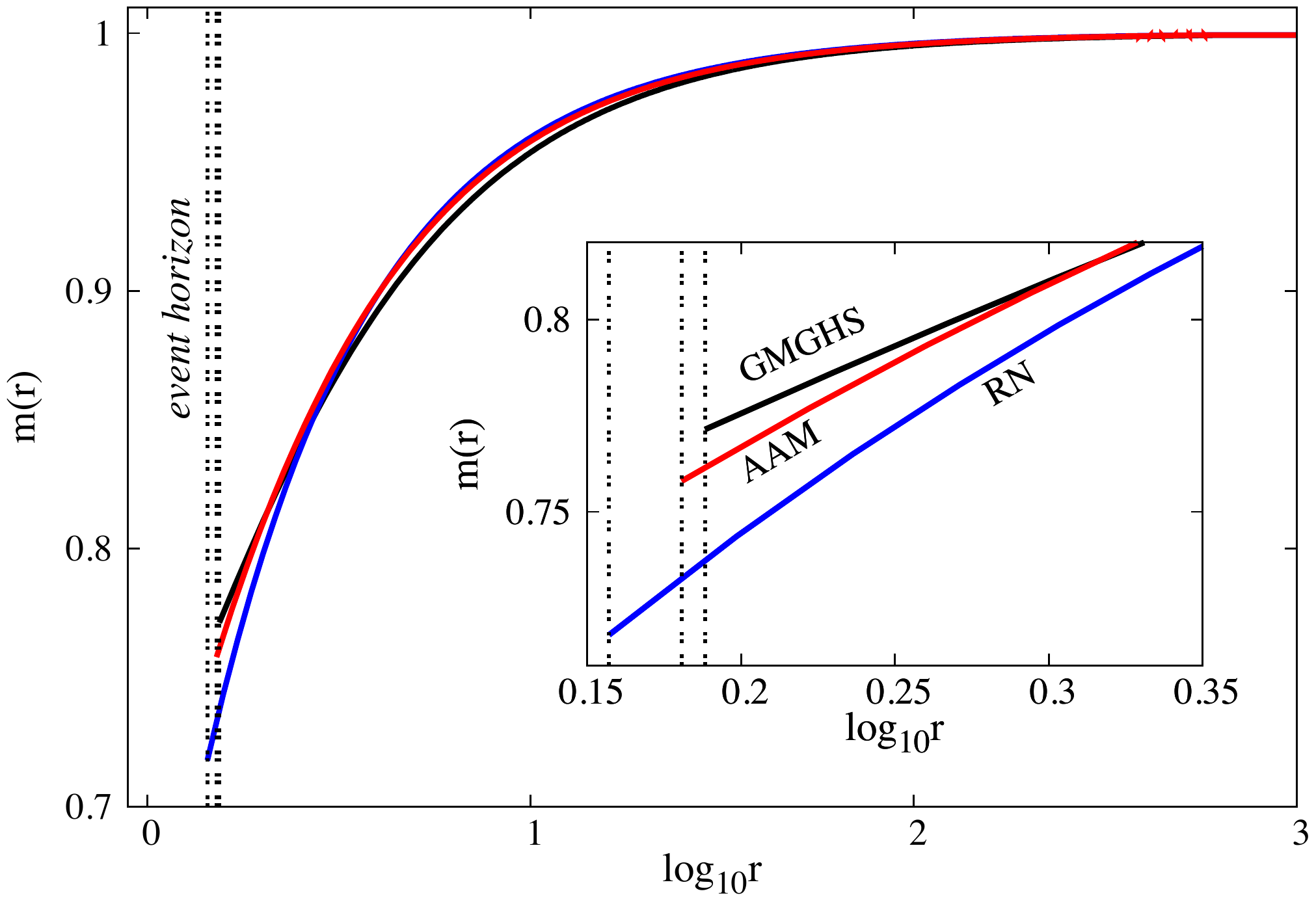}  
\includegraphics[width=0.49\linewidth,angle=0]{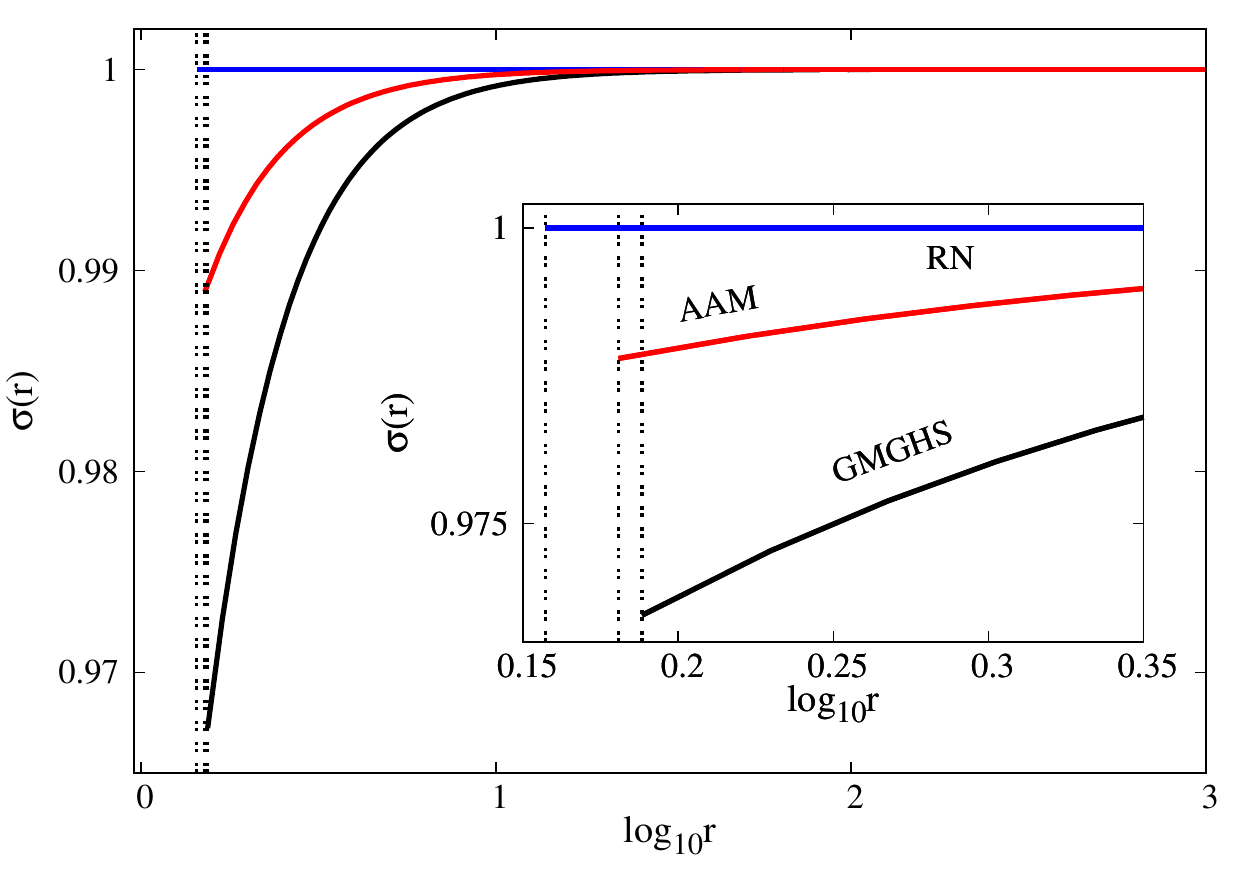}
\\
\includegraphics[width=0.49\linewidth,angle=0]{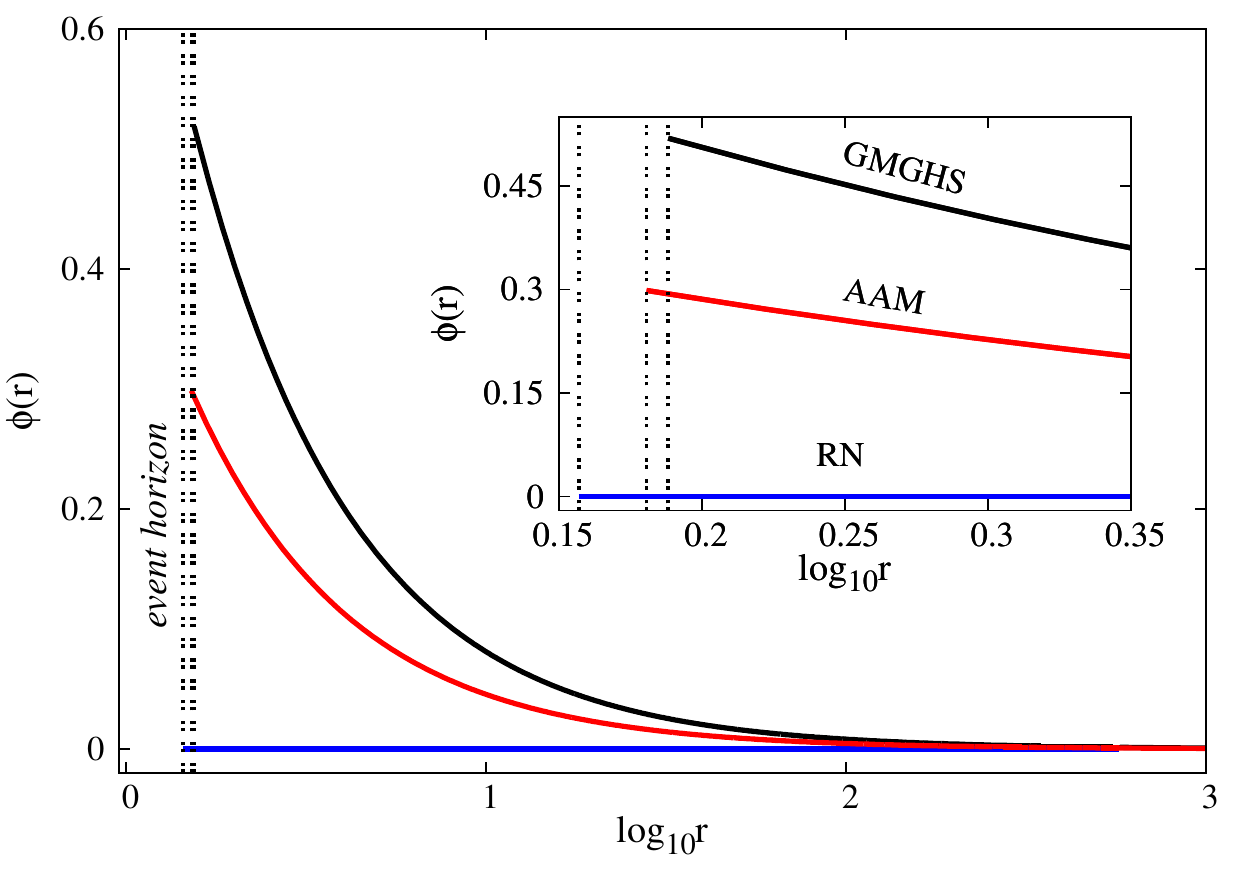}
\includegraphics[width=0.49\linewidth,angle=0]{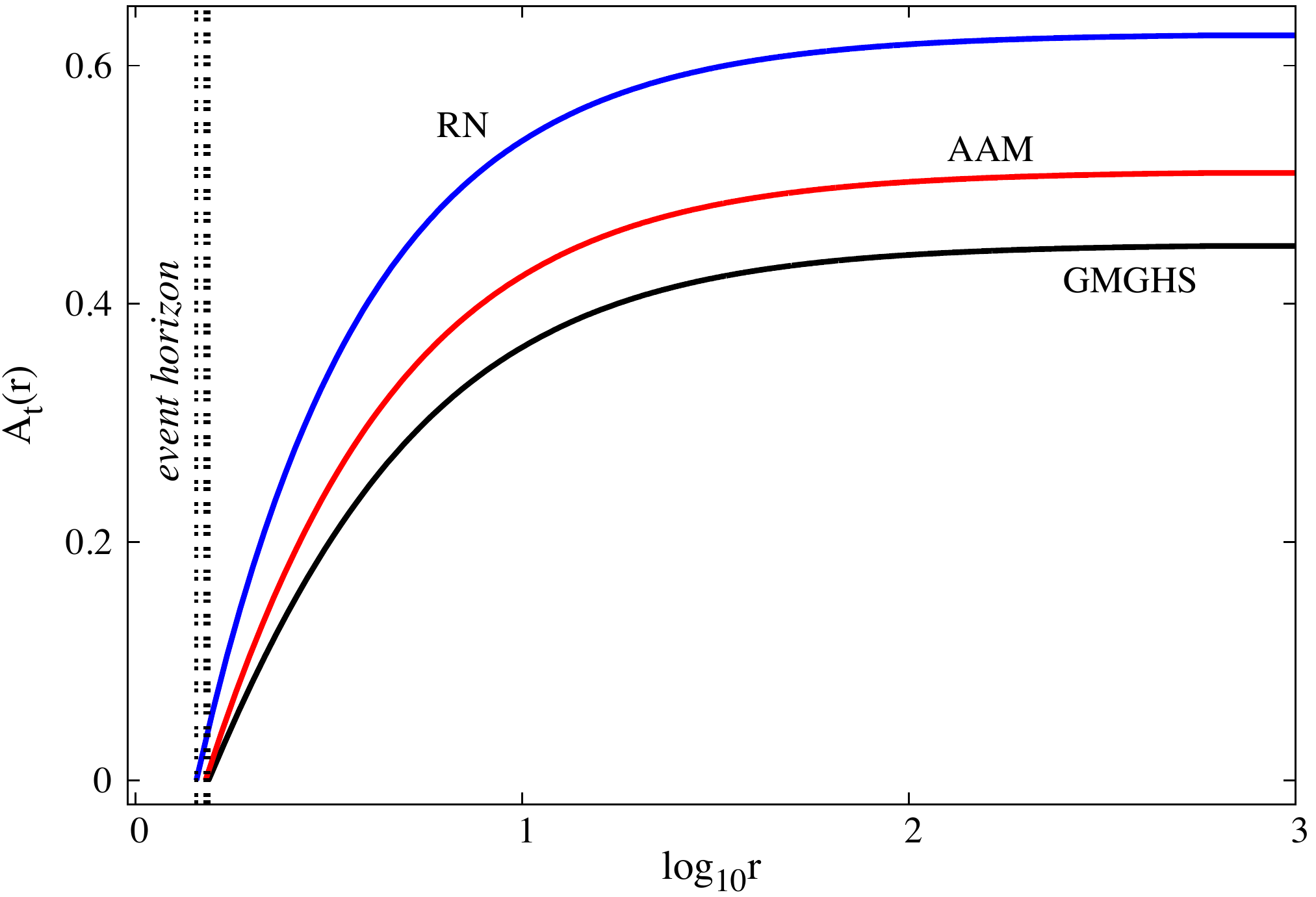}
\caption{The metric functions $m(r)$ (top left panel) and $\sigma(r)$ (top right panel), the scalar field profile $\phi(r)$ (bottom left panel) and the electric potential $A_t(r)$ (bottom right panel), for the RN BH ($\alpha=\infty$), a AAM solution  ($\alpha=100$) and a GMGHS ($\alpha=0$), all with $q=Q_e/M=0.9$.
}
\label{Fig1b}
\end{figure}

The domain of existence of these solutions 
in the $(q,a_H)$ and $(t_H,a_H)$-planes is shown in Fig.~\ref{Fig1},
where the parameter $\alpha$ spans the whole positive real line. Since $\alpha$ is dimensionful, a dimensionless parameter is obtained as $\alpha M^2$.
One can see that the solution (\ref{s1})-(\ref{s4}) 
interpolate between 
the dilatonic GMGHS BHs, occurring for $\alpha=0$ and RN BHs,
which are approached as $\alpha M^2 \to \infty$
in which case
$\phi\rightarrow 0$.  The latter limit can be shown as follows.  First define a new constant $Q$ as
\begin{equation}
Q_s=\sqrt{2Q}\left(\frac{3}{\alpha}\right)^{1/4} \ ;
\end{equation}
introducing in the solution (\ref{s1})-(\ref{s4}), the limit $\alpha \to \infty$ yields
\begin{equation}
Q_s=\mathcal{O}\left(\frac{1}{\alpha}\right)^{1/4}  \ , \qquad Q_e=Q+\mathcal{O}\left(\frac{1}{\alpha}\right)^{1/4} \ , \qquad M=\frac{Q^2+r_H^2}{2r_H}+\mathcal{O}\left(\frac{1}{\alpha}\right)^{1/4} \ .
\end{equation}
This corresponds to RN BH with charge $Q$ and horizon radius $r_H$. The metric and gauge functions can be checked to have the correct limit. In Fig.~\ref{Fig1b} we plot the profile functions defining the solution for a RN ($\alpha=\infty$), a AAM ($\alpha=100$) and a GMGHS ($\alpha=0$) BH, all with $q=0.9$. One observes, in particular, that the behaviour of the AAM solutions is ``{\it in between}" the RN and the GMGHS solutions. 
%

Setting the electric charge to zero, the AAM solutions (\ref{s1})-(\ref{s4}) trivialise. This does not, by itself, exclude the existence 
of Einstein-scalar field BH solutions of the model (\ref{action}), without electric charge.
However, one can easily prove that no such solutions exist.
The proof is based on a Bekenstein-type argument 
\cite{Herdeiro:2015waa} 
and
uses the equation for the scalar field, 
which, for (\ref{s1}) 
takes the form
\begin{eqnarray}
\label{be1}
\frac{1}{\sigma r^2}\frac{d}{dr} \left( r^2 \sigma N \phi' \right)=\frac{dV}{d\phi} \ .
\end{eqnarray}
Multiplying this equation by $\phi$, a simple manipulation yields
\begin{eqnarray}
\label{be2}
\frac{d}{dr} \left( r^2 \sigma N \phi  \phi' \right)=\sigma r^2 \left(N \phi'^2+\phi \frac{dV}{d\phi} \right) \ .
\end{eqnarray}
When integrating the above relation between $r_H$ and infinity, the left hand side vanishes
for a regular configuration, since $N(r_H)=0$ and $\phi(\infty)=0$; the integrand of the right hand side, on the other hand,
is strictly positive (since $dV/d\phi>0$).
Thus, only $\phi\equiv 0$ is possible.
The presence of an electromagnetic field results in a strictly negative extra-term in  the right hand side of 
(\ref{be2}), 
which allows circumventing this argument.

\subsection{The extremal limit and near horizon geometry}

Unlike the $\alpha=0$ case (the GMGHS solutions), the AAM BHs with $\alpha\neq 0$ possess a regular extremal limit,
with a nonzero event horizon area,  mass, and electric charge. For GMGHS solutions, the effective potential 
\cite{Goldstein:2005hq} does not have a critical point at the horizon and so the extremal solution is a naked singularity. This can be also understood from the fact that the inner horizon is singular, the Kretschmann scalar is divergent there, and so in the extremal limit the singularity is pushed to the outer horizon. In the case of AAM solution, the effective potential has a new contribution coming from the dilaton potential, yielding a regular near horizon geometry of the extremal solution. 

The extremal limit is found by  imposing the constraint $T_H=0$,
which results in an involved relation between $r_H$ and $Q_s$ (or, equivalently, $r_H$ and $Q_e$, $cf.$~(\ref{s9})).
Therefore, similarly to the RN case, the extremal AAM solutions are characterized by a single parameter, which is more conveniently taken as the event horizon radius, $r_H$. A different route to study this limit is to impose that the general model (\ref{action})
possesses an $AdS_2\times S^2$ geometry as a solution and obtain the horizon data by using the entropy function formalism \cite{Sen:2005wa, Astefanesei:2006dd,  Sen:2007qy}.\footnote{The corresponding extremal BH solution in Anti-de Sitter spacetime, for a general potential, was studied in \cite{Anabalon:2013sra}.} This configuration describes the near horizon geometry of an extremal  BH,
and has a Robinson-Bertotti--type line element 
\cite{Robinson:1959ev, Bertotti:1959pf},
with
\begin{equation}
\label{AdS2S2}
  ds^2=v_0\left(-r^2dt^2+\frac{dr^2}{r^2}\right)+v_1(d\theta^2+\sin^2\theta d\varphi^2) \ .
\end{equation}
The matter fields ansatz is
\begin{equation}
\label{matter-attractors1}
A= \tilde e\, r  \,dt \ , \qquad \phi=\phi_{\rm H} \ .
\end{equation}
Here, $v_0,v_1, \tilde e, \phi_{\rm H} $ are constants that determine the near geometry and the values of the electric field and dilaton at horizon.
Under this ansatz, the equations of the model reduce to the following three \textit{algebraic} relations
\begin{equation}
\label{matter-attractors2}
\tilde e^2= e^{-\phi_{\rm H} }\frac{v_0(v_0+v_1)}{2 v_1}\ , 
\qquad  
 \frac{1}{v_1}-\frac{1}{v_0} =V(\phi_{\rm H}) 
\ , 
\qquad  
\frac{2\tilde e^2 e^{\phi_{\rm H}}}{v_0^2}=V'(\phi_{\rm H})~.
\end{equation}
The constants $v_0$ and $v_1$ determine the $AdS_2$ `radius' and  $S^2$ radius, respectively. The radius of the horizon can be computed, as usual, from the relation $v_1=r_H^2$ and the constant $\tilde e$ determines the electric charge $Q_e$ via the conservation of the flux,
\begin{equation}
\label{matter-attractors3}
Q_e=\tilde e\frac{v_1}{v_0}e^{ \phi_{\rm H} } \ .
\end{equation}

Thus, one finds that the parameters $v_0$ and $v_1$,
which enter the near horizon geometry, as well as the electric charge $Q_e$, are fixed by the value of the scalar field at the horizon
$\phi_{\rm H} $,
yielding a continuum of solutions with
\begin{equation}
\label{matter-attractors4}
 v_0=\frac{2}{V'(\phi_{\rm H}) - V(\phi_{\rm H})} \ , ~~
 v_1=\frac{2}{V'(\phi_{\rm H})  + V(\phi_{\rm H})} \ , ~~
Q_e=\frac{e^{\phi_{\rm H}/2}\sqrt{V(\phi_{\rm H})}}{V'(\phi_{\rm H})  + V(\phi_{\rm H})} \ .
\end{equation}
Equivalently, one can obtain all horizon data as a function of the physical electric charge, but since we can not solve analytically the last equation to obtain the value of the dilaton at the horizon, we prefer to work with $\phi_H$ rather than with $Q_e$. We can explicitly check that all the expressions (\ref{matter-attractors4}) are finite at the horizon for the potential (\ref{V}).
 
Since the entropy, $S$, of the BHs with this 
near horizon geometry is given by the Bekenstein-Hawking formula,
it can be computed as $S=\pi v_1$. Consequently, (\ref{matter-attractors4}) together with (\ref{V}) imply a relatively simple expression 
for the entropy of extremal AAM solutions  as a function
of the scalar field at the horizon:\footnote{A straightforward computation using the effective potential method of \cite{Goldstein:2005hq} produces the same result.}
\begin{equation}
\label{matter-attractors5}
S=\frac{\pi}{\alpha}
\left[
2+2\phi_{\rm H}+e^{\phi_{\rm H}}(\phi_{\rm H}-2)-\sinh(\phi_{\rm H})
\right]^{-1}.
\end{equation}  
If $\alpha=0$ (the GMGHS limit) there is no regular $AdS_2\times S^2$ geometry. Following the corresponding analysis in the Section 2.2, to recover the RN limit we  first define
\begin{equation}
\label{matter-attractors6}
\phi_{\rm H}=\sqrt{\frac{2}{ Q}}
\left(\frac{3}{\alpha} \right)^{1/4}~,
\end{equation}  
with $Q$ a new constant.
Then, the limit $\alpha \to \infty$
yields
 $v_0=v_1=Q^2$
and
$S=\pi Q^2$, 
which are the results for an 
extremal RN solution with
$Q_e=Q$.

\section{Mode stability - linear analysis}
\label{sec3}
Let us now consider the dynamical stability of the AAM BHs described in section~\ref{sec2}. Following the analysis in \cite{Blazquez-Salcedo:2019nwd}, we first consider mode stability against linear perturbations of the metric and fields. We perform the analysis first for spherically symmetric perturbations and, subsequently for generic non-spherical perturbations. 

\subsection{Spherical perturbations}

Following the standard method, we consider linear spherically symmetric perturbations of the AAM BHs (see for example \cite{Blazquez-Salcedo:2018jnn}). This can be implemented using the following ansatz:
\begin{eqnarray}
\label{stab13}
ds^2=- S(r,t) dt^2+P(r,t) dr^2+r^2(d\theta^2+\sin^2 \theta d\varphi^2) \ , ~~ A=   a_0(r,t) dt \ ,~~ \phi=\phi(r,t) \ ,
\end{eqnarray}
where
\begin{eqnarray}
\label{stab3}
&&
P(r,t)=\frac{1}{N(r)}+\epsilon P_1(r)e^{-i \omega t} \ , \qquad  S(r,t)=f(r)[1+\epsilon S_1(r)e^{-i \omega t}] \ ,
\\
&&
\nonumber
 \phi(r,t)=\phi_0(r)+\epsilon \phi_1(r)e^{-i \omega t} \ , \qquad  a_0(r,t)=a_0(r)+\epsilon V_1(r)e^{-i \omega t} \ .
\end{eqnarray}
Here  
$N(r)$,
$f(r)=N(r)\sigma^2(r)$,
$\phi_0(r)$
and
$a_0(r)$
correspond  to the unperturbed solutions (\ref{s1})-(\ref{s4}); $P_1(r),S_1(r),\phi_1(r), V_1(r)$ are the perturbation functions all associated to a Fourier mode with frequency $\omega$, and $\epsilon$ is an infinitesimal parameter. The frequency is in general a complex number, $\omega=\omega_R + i\omega_I$. The real part $\omega_R$ is related with the oscillation frequency of the perturbation. If the imaginary part $\omega_I$ is negative, then the perturbation is exponentially damped with damping time $1/\omega_I$. Mode instabilities would occur if QNMs with $\omega_I>0$ exist.  

A straightforward computation shows  that both the metric and the matter fields perturbations  
are determined by $\phi_1$.
As such, the study of the system reduces to a single equation for the 
scalar field perturbation.
This equation can be written in  the standard 1D Schr\"odinger form:
\begin{eqnarray}
\label{stab5}
\left(-\frac{d^2 }{dx^2}+U_{\omega}\right) \Psi=\omega^2 \Psi \ ,
\end{eqnarray}
where we have defined the `tortoise' coordinate $x$, and the new function $\Psi$ by
\begin{eqnarray}
\label{stab6}
\frac{dx}{dr}\equiv \frac{1}{\sqrt{fN}}\  , \qquad {\rm and} \qquad  \Psi\equiv r \phi_1 \ .
\end{eqnarray}
The perturbation potential $U_{\omega}$ has the unenlightening form
\begin{equation}
\label{stab7}
U_{\omega}= f
\bigg[
V''(\phi_0)+r\phi_0' V'(\phi_0)-\frac{V(\phi_0)}{2}
+\frac{e^{-\phi_0}  Q_e^2(1-2 r\phi_0')}{r^4}
-\frac{N}{2} \phi_0'^2
\left(
1+\frac{rf'}{f}-\frac{r^2 \phi_0'^2}{4}
\right)+\frac{1-N}{r^2}
\bigg] \ .
\end{equation}
Expressed in this 1D Schr\"odinger form, the diagnosis of an unstable mode solution of (\ref{stab5}) would be $\omega^2<0$ (or more explicitly, $\omega_R=0$ and $\omega_I>0$ with $\Psi|_{r_H}=\Psi|_{\infty}=0$). Since the potential  is regular in the entire $x-$range and
 it vanishes at the BH event horizon and at infinity, this would be a bound state. A standard result in quantum mechanics is that (\ref{stab5}) will have no bound states if the potential $U_{\omega}$ is everywhere greater than the lower of its two asymptotic values.
Although the potential is not manifestly positive definite, scanning the space of solutions, this positivity is indeed satisfied. This means that $\omega^2$ is positive and the AAM BHs are stable against spherical perturbations.

\subsection{Non-spherical perturbations}

After a decomposition using tensor spherical harmonics, the generic non-spherical perturbations can be differentiated in two decoupled channels, axial perturbations and polar perturbations, depending on how they transform under reflection of the angular coordinates \cite{Kokkotas:1999bd,Nollert:1999ji,Berti:2009kk,Konoplya:2011qq}.

The axial channel only perturbs the metric and the gauge field; not the scalar field~\cite{Blazquez-Salcedo:2019nwd}.
The  ansatz for this sort of perturbations introduces three perturbation functions, $h_0$, $h_1$ and $W_2$. It reads:
\begin{eqnarray}
\label{metric_axial_pert}
&&
ds^2=-f(r) dt^2 
+ \frac{dr^2}{N(r)} +r^2(d\theta^2+\sin^2 \theta d\varphi^2)
\nonumber 
\\
&&
{~~~~~~}
- 2 \epsilon e^{-i\omega t}\bigg[h_0(r)dt+h_1(r)dr\bigg]\left[\frac{\partial_{\phi}Y_{lm}(\theta,\phi)}{\sin{\theta}} d\theta -\sin{\theta}\partial_{\theta}Y_{lm}(\theta,\phi) d\phi \right] \ ,
\\
&&
A= a_0(r) dt -\epsilon W_2(r)e^{-i\omega t}\left[ \frac{\partial_{\phi}Y_{lm}(\theta,\phi)}{\sin{\theta}}d\theta
-  \sin{\theta}{\partial_{\theta}Y_{lm}(\theta,\phi)} d\phi\right] \  ,
\end{eqnarray}
where $Y_{lm}(\theta,\phi)$ are the usual spherical harmonics. The system of equations obtained from the linearised Einstein-Maxwell-dilaton equations consists of two first order differential equations for $h_0$ and $h_1$ (the metric perturbations) coupled to a second order differential equation for $W_2$ (the electro-magnetic perturbation).

The polar channel, on the other hand, perturbs all the fields: the metric, the gauge field and the scalar field. The ansatz in this case introduces eight perturbation functions, $H_0(r)$, $H_1(r)$, $L(r)$, $T(r)$, $a_1(r)$, $W_1(r)$, $V_1(r)$ and $\phi_1(r)$. Now it reads:
\begin{eqnarray}
ds^2&=&-f(r) dt^2 
+ \frac{dr^2}{N(r)} +r^2(d\theta^2+\sin^2 \theta d\varphi^2) 
\label{metric_l0_pert}
\\
&&
- \epsilon e^{-i \omega t} Y_{lm}(\theta,\phi) \left\{  \bigg[ H_0(r) dt
+ 2  H_1(r) dr\bigg]dt+ \frac{L(r)}{N(r)}dr^2 
+2T(r)\bigg(d\theta^2+\sin^2 \theta d\varphi^2\bigg)\right\} \ ,  
\nonumber
\end{eqnarray}
\begin{eqnarray}
\label{stab2}
A&=& a_0(r)dt + \epsilon e^{-i \omega t}  \left\{Y_{lm}(\theta,\phi)  \bigg[a_{1}(r)dt +  W_1(r) dr\bigg]  + V_1(r) \bigg[ \partial_{\theta}Y_{lm}(\theta,\phi) d\theta  +  \partial_{\phi}Y_{lm}(\theta,\phi) d\phi \bigg] \right\}
\nonumber \ ,
\end{eqnarray} 
\begin{equation}
\phi=\phi_0(r) + \epsilon e^{-i \omega t} Y_{lm}(\theta,\phi)  \phi_1(r) \ .
\end{equation}
It is convenient to define
\begin{eqnarray}
\nonumber
&&
-i\omega W_1(r) - \frac{dW_1}{dr} \equiv F_0(r) \ ,
\\
&&
 -i\omega V_1(r) - a_1(r) \equiv F_1(r) \ ,
\\
\nonumber
&&
-W_1(r) + \frac{dV_1}{dr} \equiv F_2(r) \ ,
\end{eqnarray}
 Introducing this ansatz on the field equations and performing a number of algebraic manipulations, it is possible to see that the polar perturbations are described by: two first order differential equations for $H_1$ and $T$ (metric perturbations), another two first order differential equations for $F_0$ and $F_1$ (electro-magnetic perturbations), and a second order differential equation for $\phi_1$ (scalar perturbation).  These six functions determine the rest of functions ($H_0$, $L$ and $F_2$) via some algebraic equations.

As a summary, the minimal system of equations for the non-spherical perturbations can be written like~\cite{Blazquez-Salcedo:2019nwd}
\begin{eqnarray}
\partial_r \Psi_{j} = M_{j} \Psi_{j} \ ,
\end{eqnarray}
where $j=\{\mathit{Axial}, \mathit{Polar}\}$ and we have the perturbation functions for each of these channels:
\begin{eqnarray}
\Psi_{\mathit{Axial}} = \left[ h_0, h_1;  W_2, \partial_r W_2 \right] \nonumber \ , \\
\Psi_{\mathit{Polar}} = \left[ H_1, T;  F_0, F_1; \phi_1, \partial_r \phi_1 \right] \nonumber \ .
\end{eqnarray}
We have separated with semicolons the space-time perturbations, electromagnetic perturbations and scalar perturbations (only present in the polar channel).
The coefficients of the $4\times 4$ matrix $M_{\mathit{Axial}}$ and the $6\times 6$ matrix $M_{\mathit{Polar}}$ depend on the unperturbed metric and field functions, the $l$ harmonic index and the mode complex frequency $\omega$. 
\subsection{The quasinormal modes}
The AAM BH are solutions of the model (\ref{action}), with $\gamma=1$ and arbitrary $\alpha$, and as we have seen these solutions are known in closed form, $cf.$ section~\ref{sec2}. For the analysis of linear stability and QNMs spectrum, however, we have obtained numerically the BHs in the more general case of model (\ref{action}) with both values of $\gamma$ and $\alpha$ arbitrary. This allows us to better explore limiting cases. For example, varying the dilaton coupling $\gamma$ from $\gamma=1$ to $\gamma=0$,  provides a simple procedure to recover the electrovacuum model. In this way, we can continuously track the QNMs, connecting them to all known spectra:  Schwarzschild, RN and GMGHS \cite{Ferrari:2000ep}. 
This is an excellent cross-check on the numerical calculation of QNMs.

Following this methodology and the procedure described in~\cite{Blazquez-Salcedo:2019nwd} (see also \cite{Blazquez-Salcedo:2018pxo}), we have computed the spectrum of QNMs of the AAM BHs. When the model reduces to electrovacuum $\gamma=0=\alpha$, our results for the QNM of the RN BH reproduce the results in~\cite{Leaver:1990zz}. In the other well known limit when $\alpha=0$, our results reproduce the spectrum calculated in~\cite{Ferrari:2000ep} for the GMGHS BHs. In these two limiting cases all the modes are stable. 

After benchmarking the method with these two limiting cases, we have tackled the AAM BHs in~\cite{Astefanesei:2019mds}, scanning for possible unstable modes, in particular for the thermodynamically stable solutions. In the illustrative case $\alpha=1$, $r_H=2$, we have scanned for unstable modes several thermodynamically stable solutions between $Q_e=3.305929$ (extremal), and $Q_e=3.11784$ (critical, separating thermodynamically stable from unstable solutions - $cf.$ Section~\ref{sec5}), as well as solutions close to the critical point ($Q_e=3, 3.1$). No unstable modes were found for the $l=1,2$ cases (note $l=0$ corresponds to the previous spherically symmetric perturbations). It is unlikely that unstable modes may exist for larger values of $l$.
\begin{figure}[t!]
\centering
\includegraphics[width=0.37\linewidth,angle=-90]{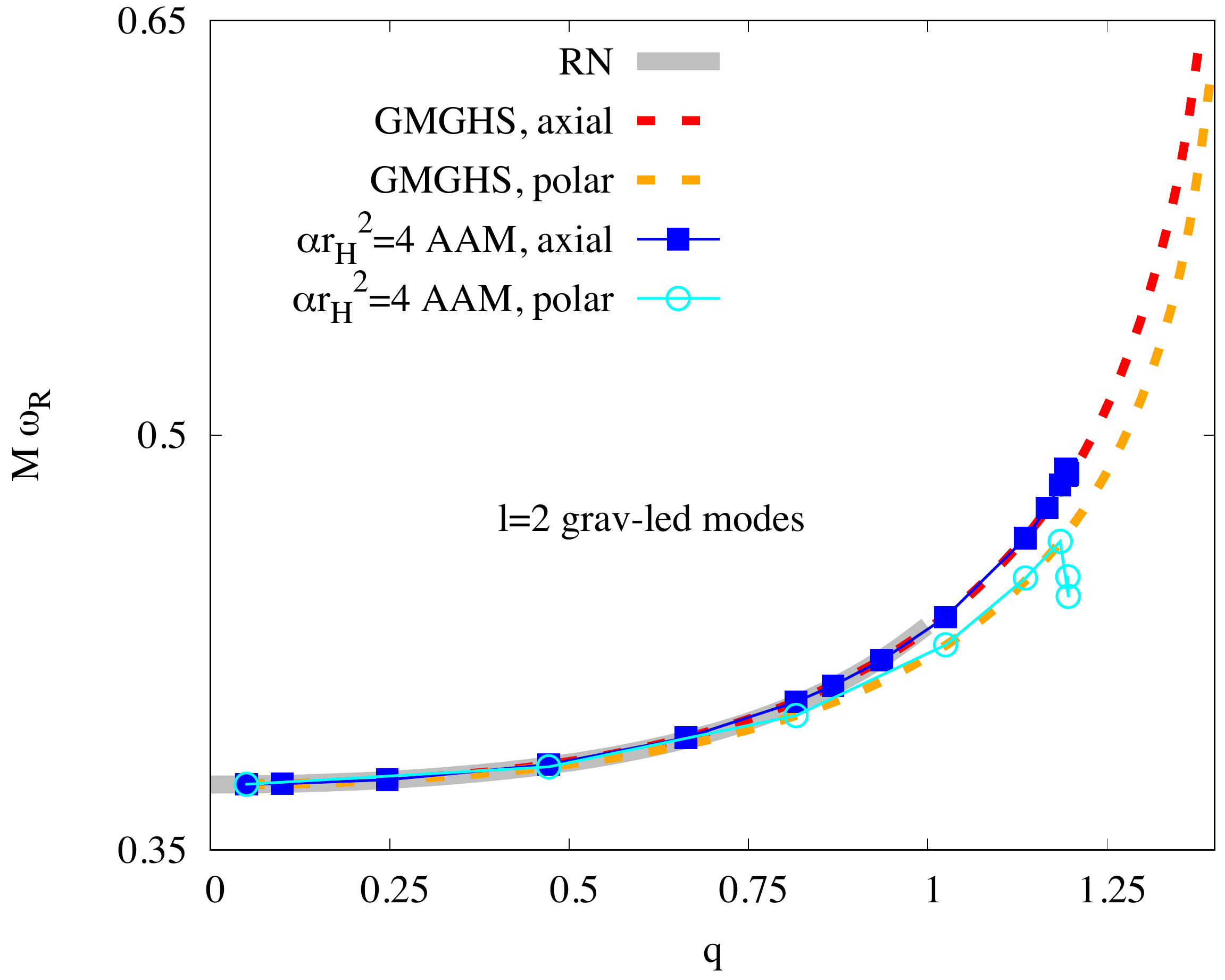} \ \ 
\includegraphics[width=0.37\linewidth,angle=-90]{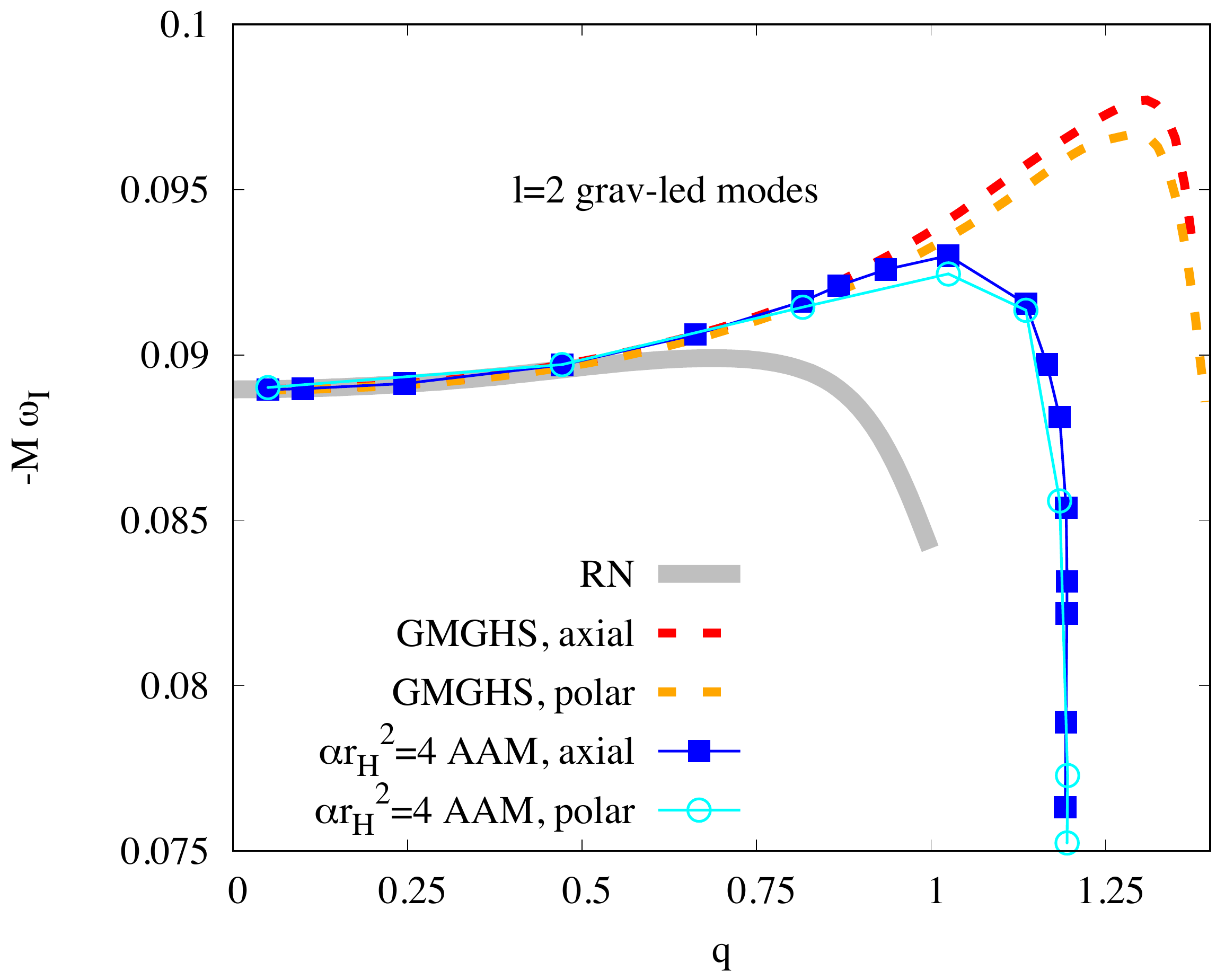}
\includegraphics[width=0.37\linewidth,angle=-90]{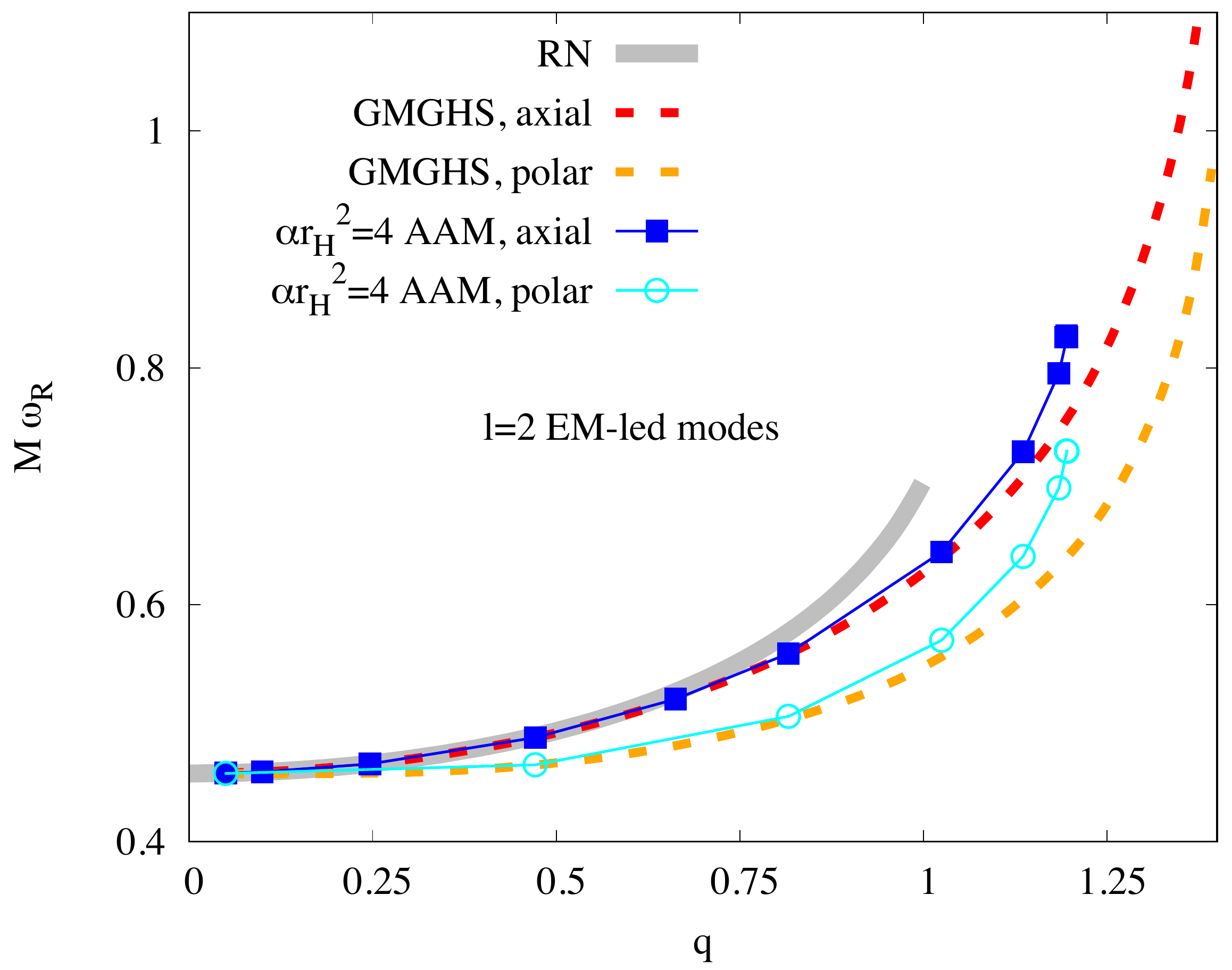} \ \ 
\includegraphics[width=0.37\linewidth,angle=-90]{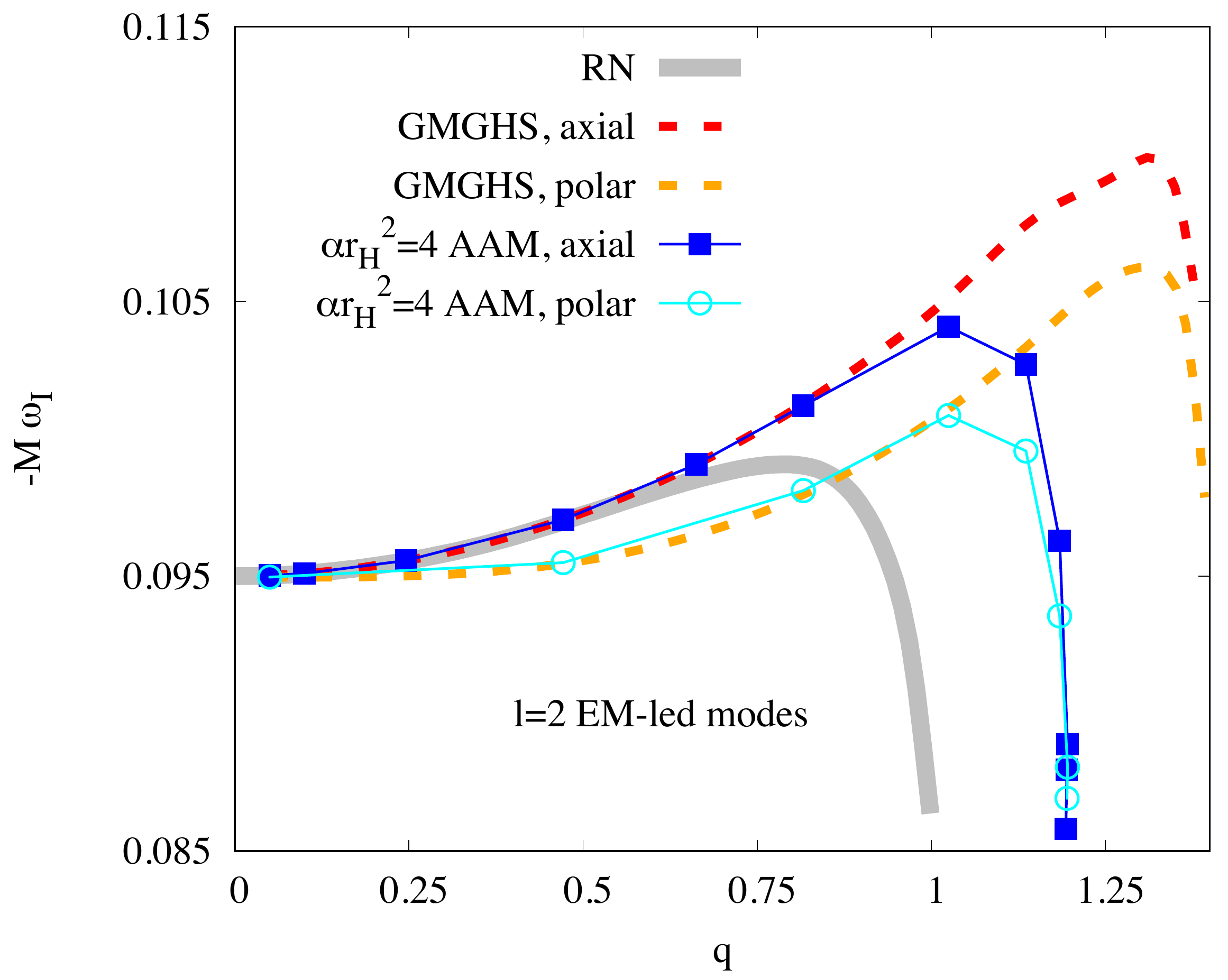} 
\includegraphics[width=0.37\linewidth,angle=-90]{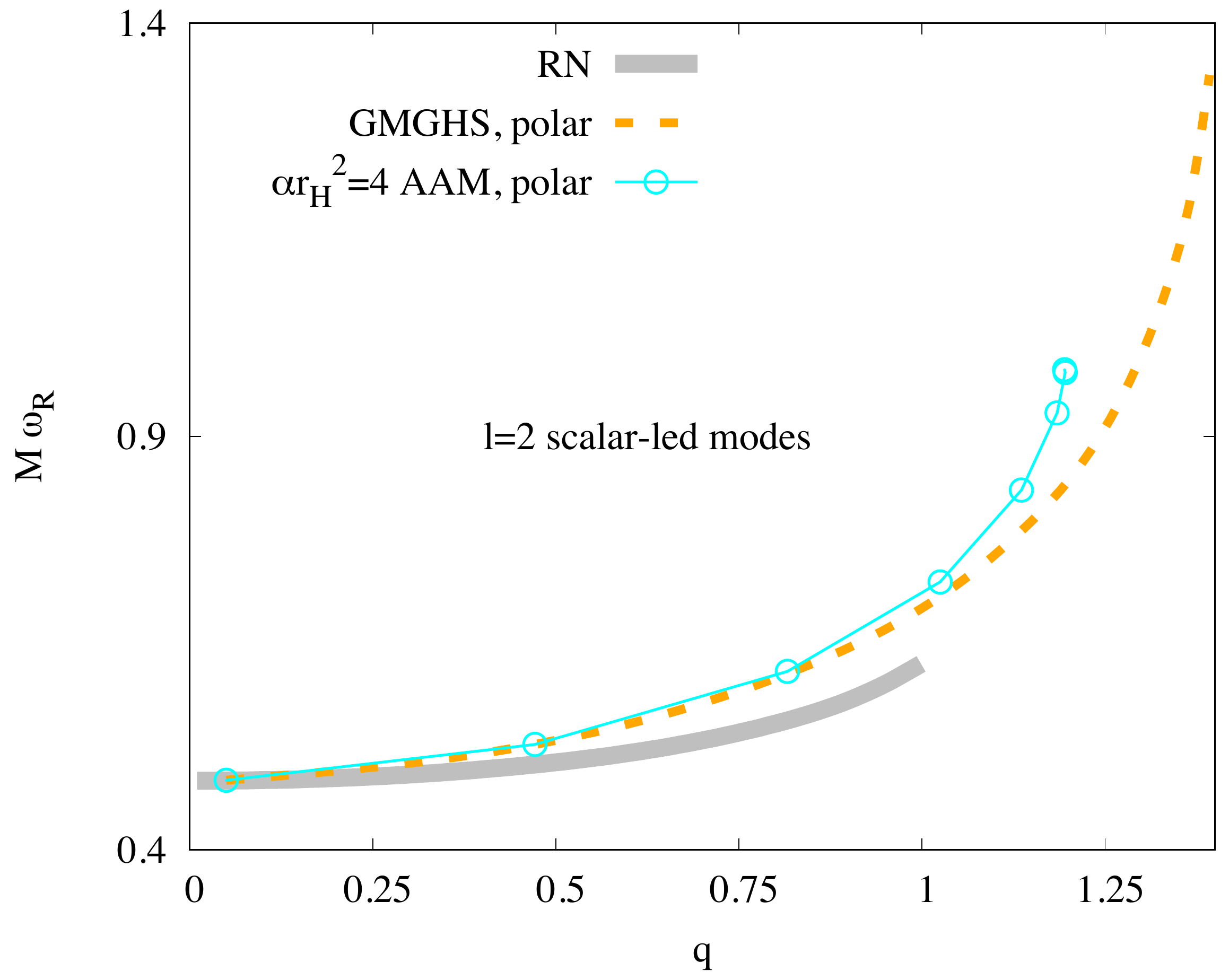} \ \ 
\includegraphics[width=0.37\linewidth,angle=-90]{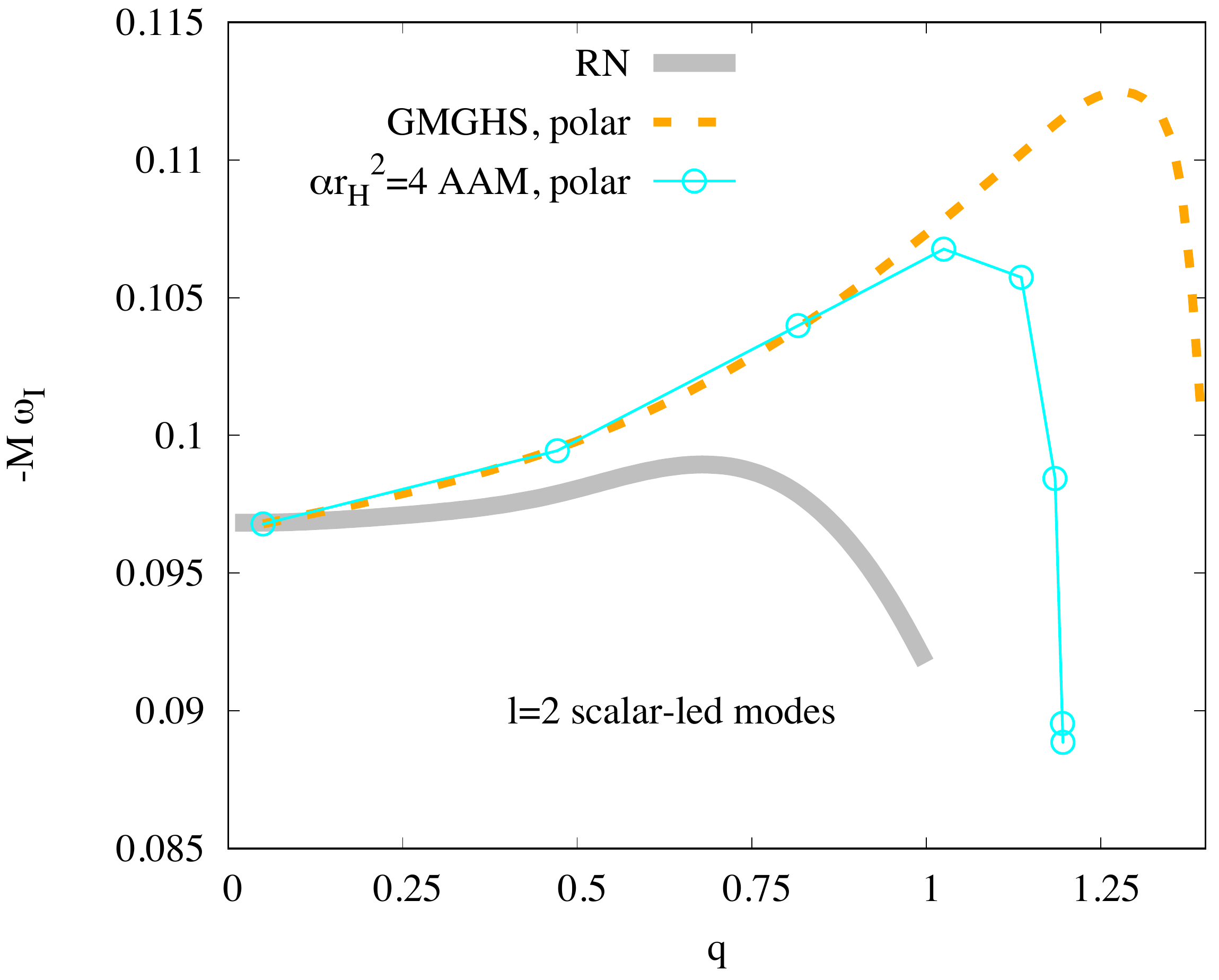} 
\caption{Real (imaginary) part of $\omega$ for $l=2$ perturbations as a function of $q$ are shown in the left (right) panels. In the top we show grav-led modes (axial and polar), in the middle EM-led modes (axial and polar), and in the bottom scalar-led modes (polar).
We show results for RN (with a continuous grey line), GMGHS (with dashed red/orange line for axial/polar modes) and the AAM BHs with $\alpha r_H^2=4$ (with dotted blue/cyan curves for axial/polar modes).}
\label{Fig2}
\end{figure}

In addition, by slowly varying $\alpha$ and $\gamma$, we have tracked the modes that for $\gamma=0=\alpha$ correspond to the RN spectrum and for $\gamma=1$, $\alpha=0$ correspond to the GMGHS spectrum. It is possible to see that all of these modes remain stable when getting to the AAM family with $\gamma=1$ and (say) $\alpha r_H^2=4$.  This is shown in Fig.~\ref{Fig2}, where the left (right) panels exhibits the scaled real (imaginary) part of $\omega$ versus $q$, for gravitational (top), electromagnetic (middle) and scalar (bottom) perturbations.  All the curves in these plots are for $l=2$ modes. With a solid thick grey line we show the RN modes. The GMGHS modes are shown with a dashed red curve (axial) and a dashed orange curve (polar). The AAM modes are shown with a dotted blue curve (axial) and a dotted cyan curve (polar).

In Fig.~\ref{Fig2} we call grav-led modes to those that correspond to purely gravitational modes when the couplings are turn to zero (Schwarzschild). Typically, for small and intermediate values of $q$, the grav-led modes excite more strongly the gravitational perturbations; that is, the amplitude of the metric perturbation functions is larger than the amplitude of the electromagnetic and scalar perturbations. Similarly, EM-led modes and scalar-led modes correspond respectively to purely electromagnetic an scalar modes in the Schwarzschild limit~\cite{Blazquez-Salcedo:2019nwd}.

In Fig.~\ref{Fig2} we can observe how the GMGHS and AAM modes diverge from the RN modes as we increase $q$. In all cases, the AAM curves fit ``in between" the RN and GMGHS curves. 
We can also appreciate how the spacetime and electromagnetic modes split into two (axial and polar) when the charge is increased, for both the GMGHS and AAM cases. For the scalar modes, no such splitting occurs,  since they only exist as polar modes. This is related with the breaking of isospectrality in these dilatonic BHs.
 It is well known that  RN modes possess isospectrality, meaning axial and polar modes are equal. This isospectrality is typically broken by the dilaton, as already noted in \cite{Ferrari:2000ep,Blazquez-Salcedo:2019nwd} (see also \cite{Blazquez-Salcedo:2016enn,Blazquez-Salcedo:2017txk}). 
 
 Moreover, the presence of the dilaton couples the scalar perturbations to the full polar equations. Thus, while the axial channel retains two families of modes -- spacetime (Fig.~\ref{Fig2} top) and electromagnetic modes (Fig.~\ref{Fig2} middle), as for RN --, the polar channel acquires three families of modes -- the previous two plus the scalar modes (Fig.~\ref{Fig2} bottom). In the limit $q=0$, all cases reduce to the Schwarzschild BH, and the modes converge to the Schwarzschild spectrum. Isospectrality is restored, and polar modes merge with axial modes of the same family. The dilatonic modes converge to the modes of a minimally coupled scalar field on the Schwarzschild background. Moreover, considering the two limits of the domain of existence shown in Fig.~\ref{Fig1}, in the limit $\alpha=0$ with $\gamma=1$, the modes converge to the QNMs of the stringy GMGHS BHs, where isospectrality is also broken. In the limit $\alpha \to \infty$,  the modes tend to converge to the RN modes, and again isospectrality is restored. 
 
 In this scanning, we have observed that all modes remain stable, although close to extremality there is a tendency to increase the damping time, corresponding to a smaller values of the imaginary part of $\omega$. This trend, however, occurs already for RN BHs.

To conclude, the spectrum of QNMs of the AAM BHs is qualitatively very similar to the one studied recently for scalarised RN BHs in \cite{Blazquez-Salcedo:2019nwd}: all QNMs are damped, although the spectrum is richer due to the broken isospectrality and the non-trivial scalar degree of freedom. All these features  strongly indicate that the AAM BHs are mode stable for arbitrary values of $\alpha$ and $q$, in both the axial and polar channels, and for arbitrary $l$ numbers.

\section{Dynamical stability - non-linear analysis}
\label{sec4}
The linear stability analysis, as discussed in Section~\ref{sec3}, does not rule out that large perturbations can cause instabilities. One way to assess this possibility is by performing fully non linear numerical simulations, within the framework of numerical relativity, starting with a highly perturbed configuration. With this goal in mind, we  have performed non-linear evolutions of the model (\ref{action}) with $\gamma=1$. 

To test the stability of an AAM BH against large perturbations we consider two scenarios. Firstly, we have started with a RN BH with some small scalar field profile around it. We face this as a highly perturbed AAM BH. This initial data has the advantage of being readily accessible within the framework of numerical relativity. It is, however, constraint violating initial data, in the sense that it does not solve the constraint equations obtained from~(\ref{action}). Nonetheless, as it often happens with constraint violating data, the evolution converges to a true, stable, solution of the model, which in this case is an AAM BH. The latter, however, has $q<1$, similarly to  the initial RN configuration. Thus, although such evolutions confirm the stability of the AAM solutions, one cannot, in this way, probe the overcharged regime, which is the most interesting one if one is interested in the thermodynamically stable BHs.  This issue can be solved in our second scenario, where we start with a GMGHS BH configuration as initial data, but in a model with $\alpha\neq 0$. Again, this is constraint violating initial data. Unlike the RN case, however, there are GMGHS BHs with $q>1$; thus, we can start with such overcharged configurations and the evolutions confirm an overcharged AAM configurations forms. In this way we show that even overcharged AAM solutions are stable in a non-linear sense. We remark, however, that the simulations we performed did not form AAM solutions precisely in the thermodynamically stable region. Reaching this region is numerically challenging within our approach, as it is very close to extremality. Nonetheless, the results herein establish that, in the sense we discussed, both undercharged and overcharge AAM solutions are non-linearly preferred. We expect this extends to the thermodynamically stable region.

Accordingly, in the first scenario we take as initial data a RN BH configuration with charge to mass ratio $q=0.2$, with the following dilaton field initial Gaussian distribution
\begin{equation} \phi=A_0e^{-(r-r_0)^2/\lambda^2} \ ;
  \end{equation} 
as an illustrative example, we have taken  $A_0=3\times 10^{-4}$, $r_0=10M$ and $\lambda=\sqrt{8}$.  We have evolved this system for the coupling $\gamma=1$ and taken $\alpha=0.01$.

The framework for this numerical evolutions is the 3+1 spacetime decomposition. For the metric, this split is given by
\begin{equation}
ds^2=-(\alpha_0^2+\beta^r \beta_r)dt^2+2\beta_r dtdr+e^{4\chi}\left[a\, dr^2+ b\, r^2 d\Omega_2\right],
\end{equation}
where the lapse $\alpha_0$, shift component $\beta^r$, and the (spatial) metric functions, $\chi,a,b$ depend on $t,r$. A conformally flat metric with $a=b=1$ is chosen together with a time symmetry condition, $i.e.$ vanishing extrinsic curvature, $K_{ij}=0$. A description of the code to perform the evolutions and previous numerical studies of dynamical scalarisation of RN BHs can be found in~\cite{Sanchis-Gual:2015lje,Sanchis-Gual:2016tcm,Hirschmann:2017psw,Herdeiro:2018wub,Fernandes:2019rez}.
The evolution are performed in spherical coordinates under the assumption of spherical symmetry. 
The time integration uses a second-order Partially Implicit Runge-Kutta (PIRK) method developed by~\cite{CorderoCarrion:2012ic,Cordero2}.

\begin{figure}[t]
\includegraphics[width=0.49\linewidth,angle=0]{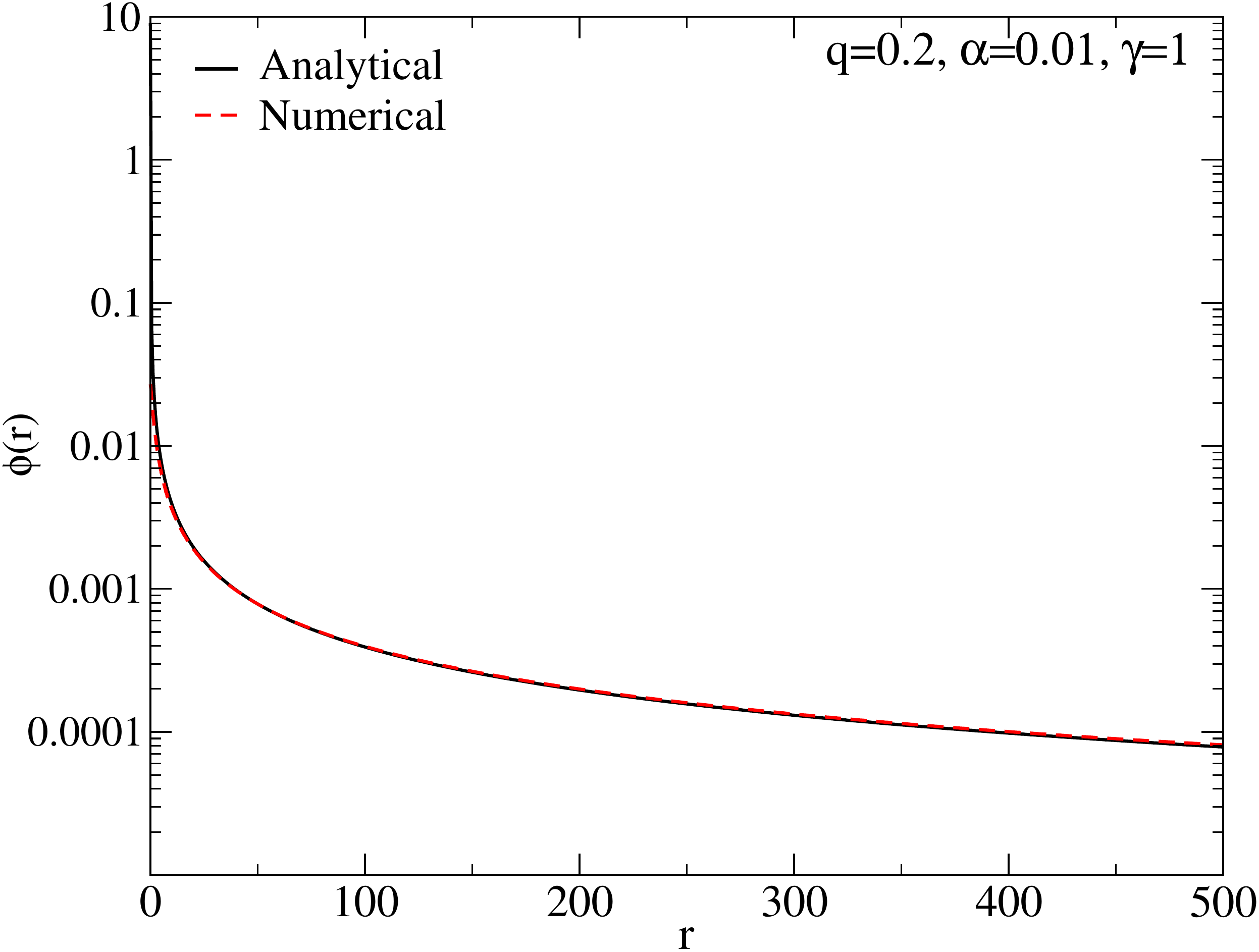}
\includegraphics[width=0.49\linewidth,angle=0]{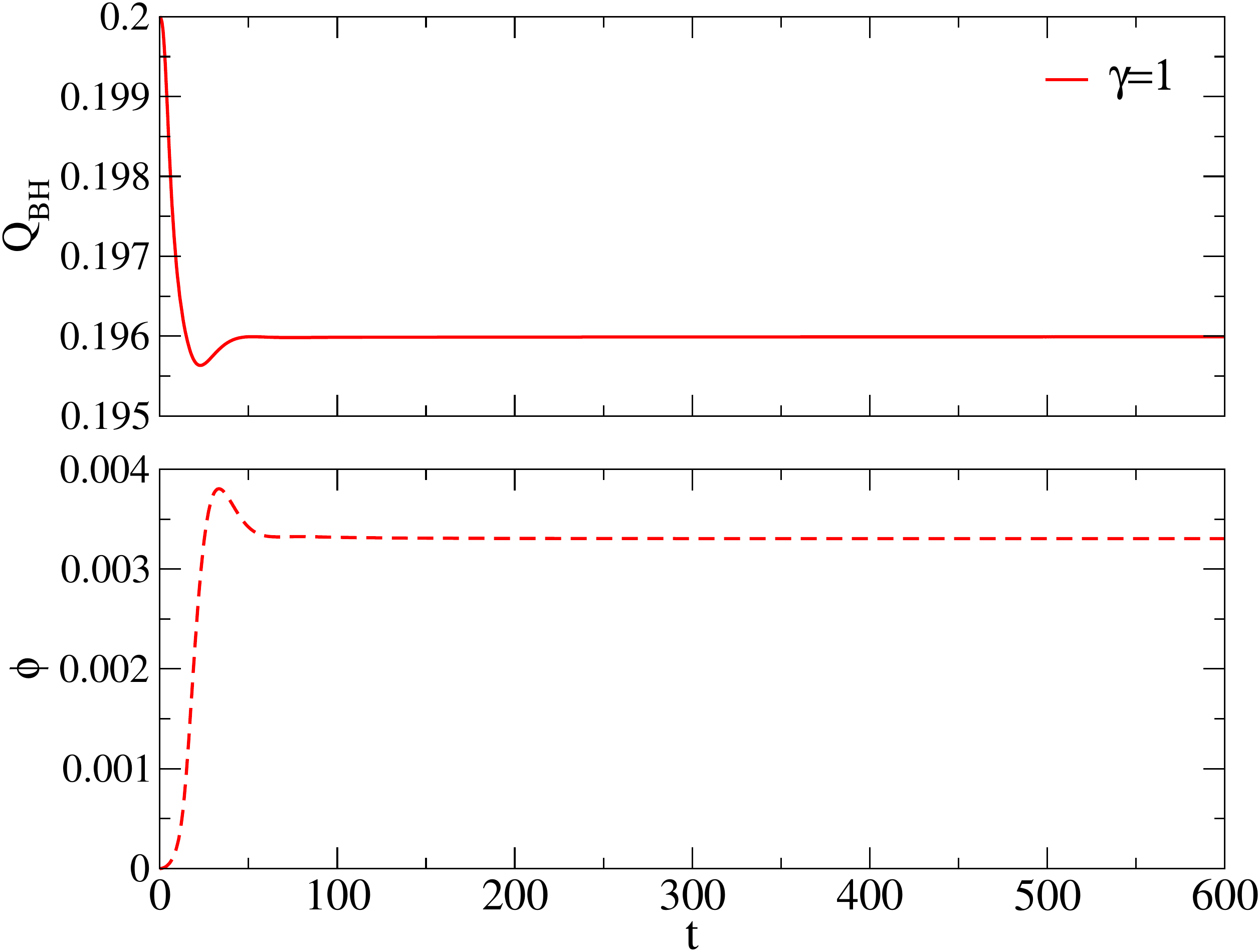}
\\
\caption{
(Left panel) Radial profile of the scalar field $\phi(r)$ at $t=600$ for the initial data consisting of a RN BH with a small dilaton Gaussian profile. There is an excellent matching between the profile obtained from the numerical evolution and the profile of a AAM BH with the same $q$ and $r_H$. 
(Right panels) Time evolution of the electric charge at the horizon $Q_{e}$ and the amplitude of the scalar field extracted at $r_0 =  11.09$.
}
\label{fig4}
\end{figure}

Let us briefly describe the evolution equations for the models (\ref{action}) that are used in our numerical evolutions, under the 3+1 split and assuming spherical symmetry. For the electric field $E^{r}$ and an extra variable, $\Psi$, introduced to damp dynamically the constrains, they take the form
\begin{eqnarray}
\partial_{t}E^{r}&=&\beta^{r}\partial_{r}E^{r}-E^{r}\partial_{r}\beta^{r}+(\alpha_0 K E^{r} -D^{r}\Psi)+\gamma\alpha_0\,\Pi E^{r}\,, \nonumber \\
\partial_{t}\Psi&=&\alpha_0(-\gamma  D_{r}\phi E^{r} - D_{i}E^{i}-\kappa_{1}\Psi)\ ,
\end{eqnarray}
where $K$ is the trace of  $K_{ij}$, we have taken $\kappa_1=1$ and $\Pi\equiv -n^{a}\nabla_{a}\phi$.

The Klein-Gordon equation is given by: 
\begin{eqnarray}
\partial_{t}\phi&=&\beta^{r}\partial_{r}\phi-\alpha_0\Pi \  , \nonumber \\
\partial_{t}\Pi&=&\beta^{r}\partial_{r}\Pi+\alpha_0 K\Pi-\frac{\alpha_0}{ae^{4\chi}}\biggl[\partial_{rr}\phi\nonumber\\
&+&\partial_{r}\phi\biggl(\frac{2}{r}-\frac{\partial_{r}a}{2a}+\frac{\partial_{r}b}{b}+2\partial_{r}\chi\biggl)\biggl]\nonumber\\
&-&\frac{\partial_{r}\phi}{ae^{4\chi}}\,\partial_{r}\alpha_0-2\gamma\alpha_0\, e^{\gamma\phi}\,a\,e^{4\chi}(E^{r})^{2} + \alpha_0 \frac{dV(\phi)}{d\phi} \ .
\label{eq:sist-KG}
\end{eqnarray}

The matter source terms  for the scalar field, to be used in the Einstein equations, read
\begin{eqnarray}
\mathcal{E}^{\rm{SF}}&\equiv &n^{\alpha}n^{\beta}T^{\rm{SF}}_{\alpha\beta}=\frac{1}{32\pi}\biggl(\Pi^{2}+\frac{\partial_{r}\phi^{2}}{ae^{4\chi}}\biggl) + \frac{1}{16\pi}V(\phi) \nonumber \\
j_{r}^{\rm{SF}}&\equiv &-\gamma^{\alpha}_{r}n^{\beta}T^{\rm{SF}}_{\alpha\beta}=-\frac{1}{16\pi}\Pi\,\partial_{r}\phi\ ,\nonumber \\
S_{a}^{\rm{SF}}&\equiv &(T^{r}_{r})^{\rm{SF}}=\frac{1}{32\pi}\biggl(\Pi^{2}+\frac{\partial_{r}\phi^{2}}{ae^{4\chi}}\biggl) - \frac{1}{16\pi}V(\phi)\ , \nonumber \\
S_{b}^{\rm{SF}}&\equiv &(T^{\theta}_{\theta})^{\rm{SF}}=\frac{1}{32\pi}\biggl(\Pi^{2}-\frac{\partial_{r}\phi^{2}}{ae^{4\chi}}\biggl) - \frac{1}{16\pi}V(\phi)\ .
\end{eqnarray}
and for the electric field
\begin{eqnarray}
\mathcal{E}^{\rm{em}}=-S_{a}^{\rm{em}}=S_{b}^{\rm{em}}=\frac{1}{8\pi}\,a\,e^{4\chi}(E^{r})^{2}e^{\gamma\phi}\ .
\end{eqnarray}
The momentum density $j_{r}^{\rm{em}}$ vanishes because there is no magnetic field in spherical symmetry.

Numerical evolutions are made under a spacetime discretisation. The logarithmic numerical grid extends from the origin to $r=1500M$ and uses a maximum resolution of $\Delta r=0.0125M$.

\begin{figure}
\includegraphics[scale=0.19,trim=0.1cm 0cm 0cm 0cm]{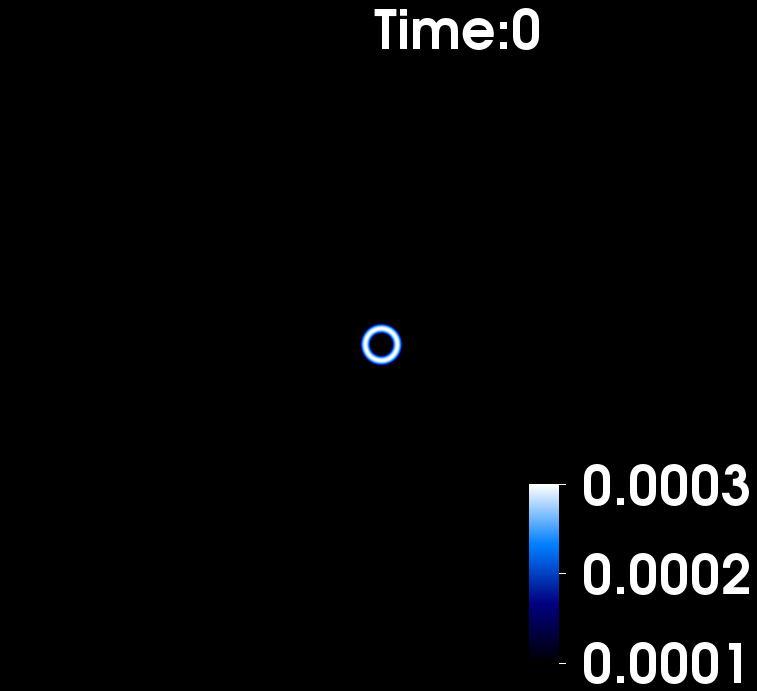}
\includegraphics[scale=0.19,trim=0.1cm 0cm 0cm 0cm]{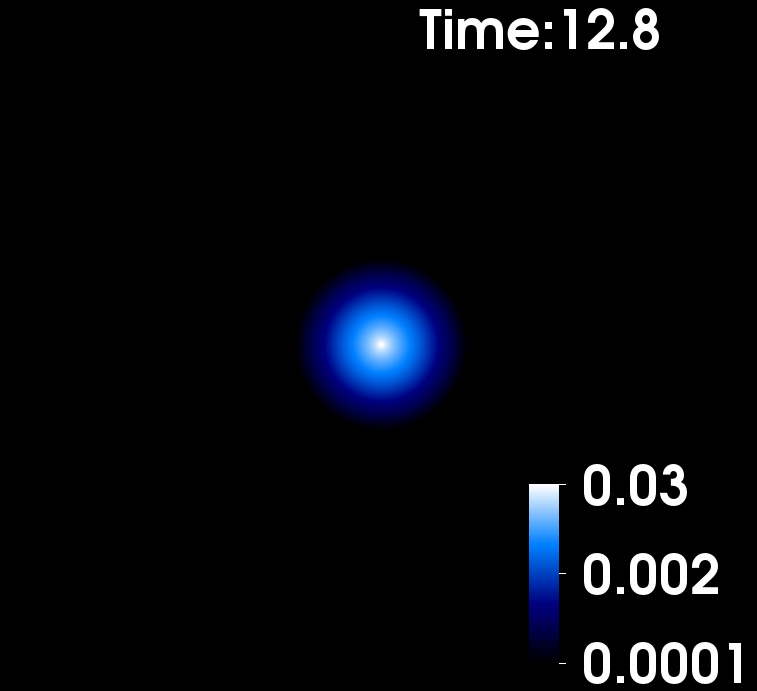}
\includegraphics[scale=0.19,trim=0.1cm 0cm 0cm 0cm]{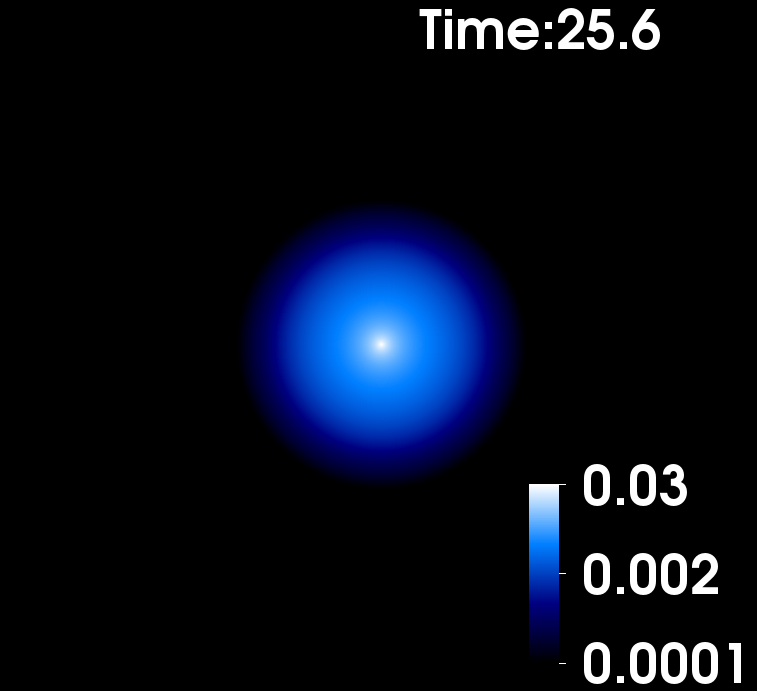}\\
\includegraphics[scale=0.19,trim=0.1cm 0cm 0cm 0cm]{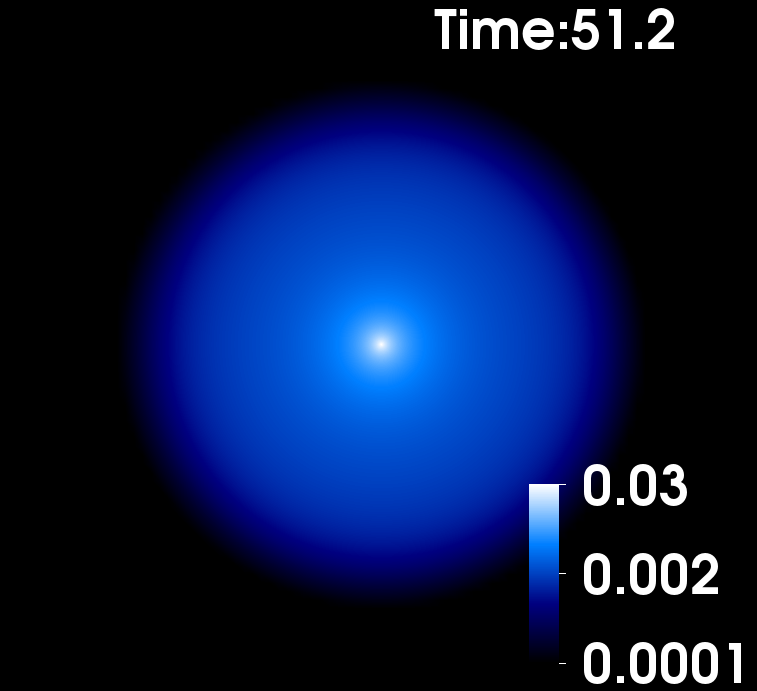}
\includegraphics[scale=0.19,trim=0.1cm 0cm 0cm 0cm]{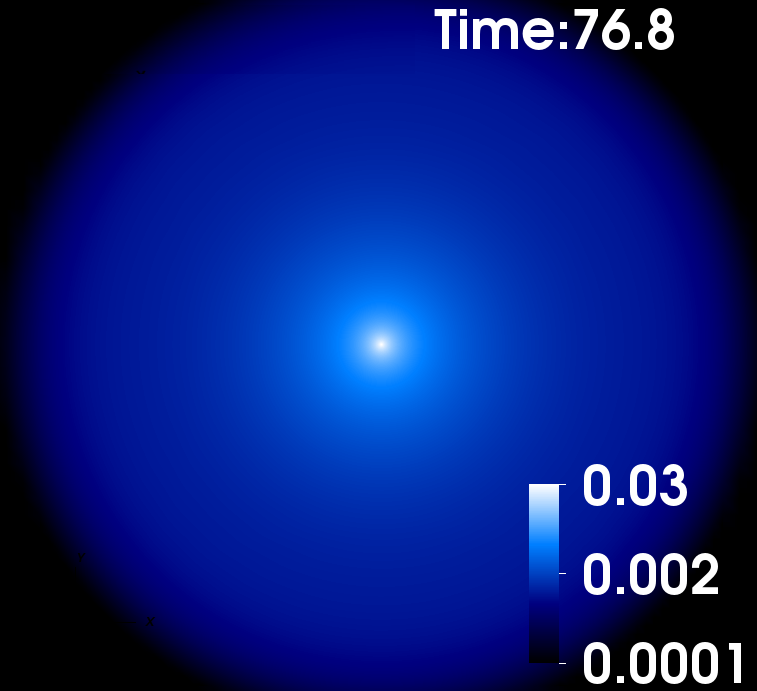}
\includegraphics[scale=0.19,trim=0.1cm 0cm 0cm 0cm]{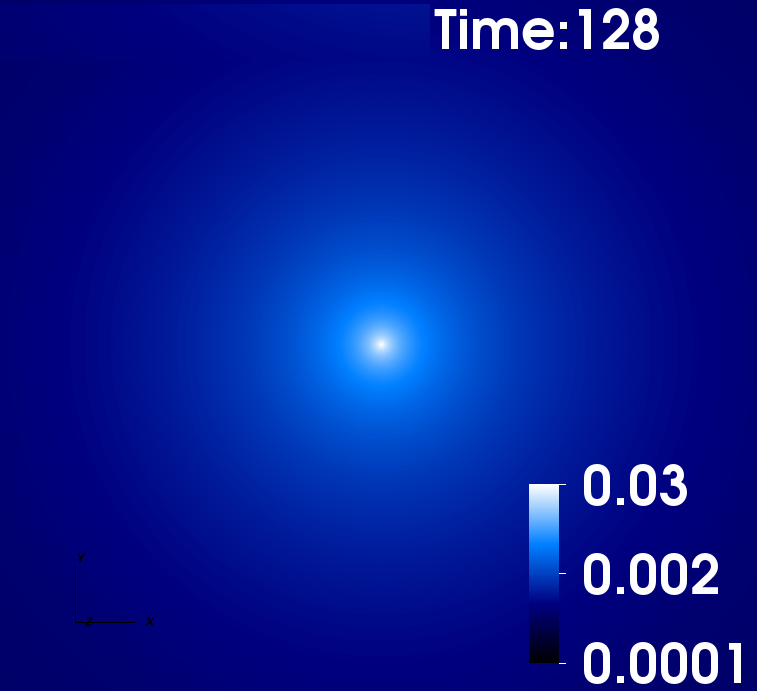}
\caption{ Snapshots of the time evolution of the scalar field profile on the equatorial plane in the evolutions that do not impose spherical symmetry. One observes the initial RN configuration with a small dilaton perturbation evolving towards a dilatonic AAM BH.}
\label{fig4n}
\end{figure}

 \begin{figure}[t]
\includegraphics[width=0.49\linewidth,angle=0]{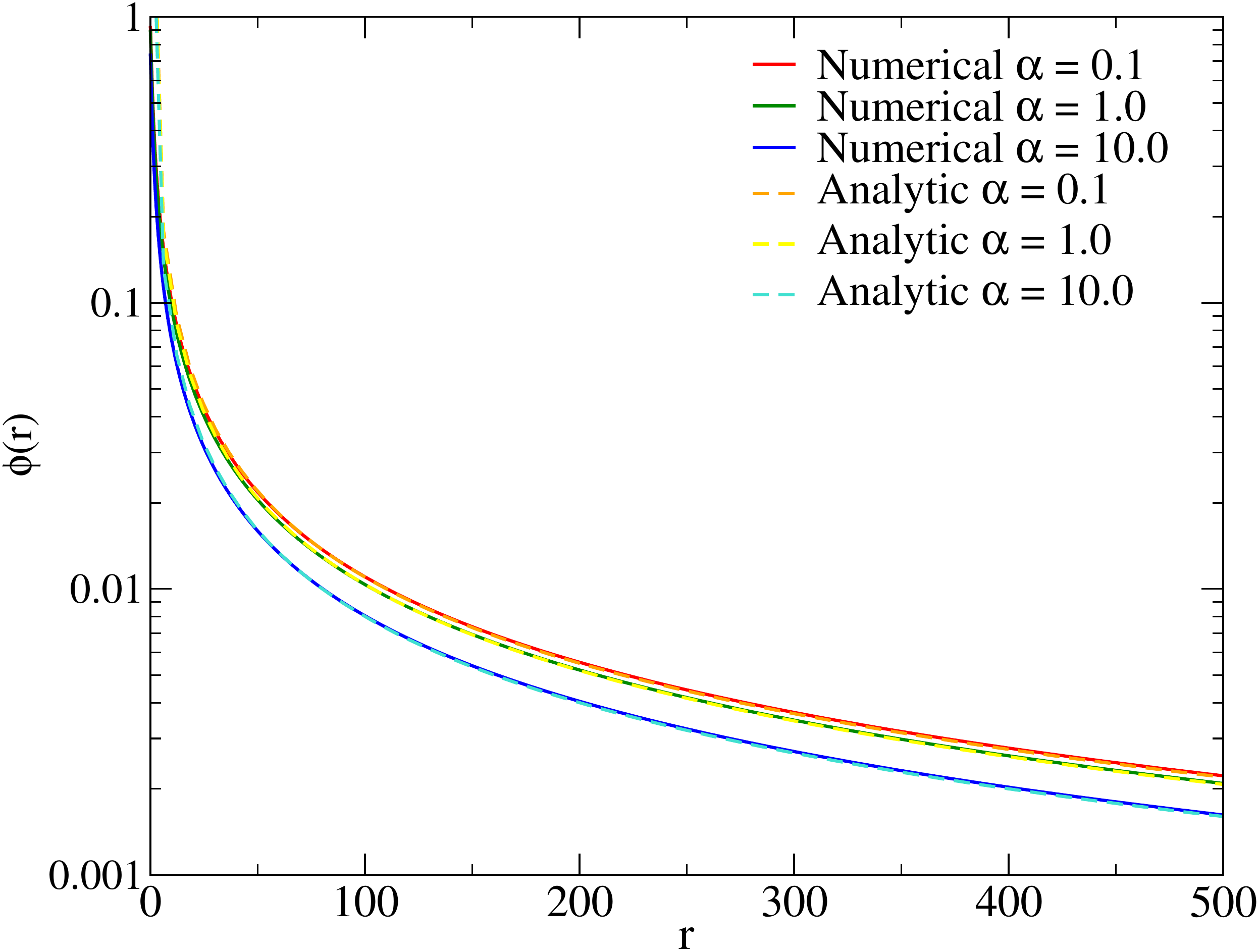}
\includegraphics[width=0.49\linewidth,angle=0]{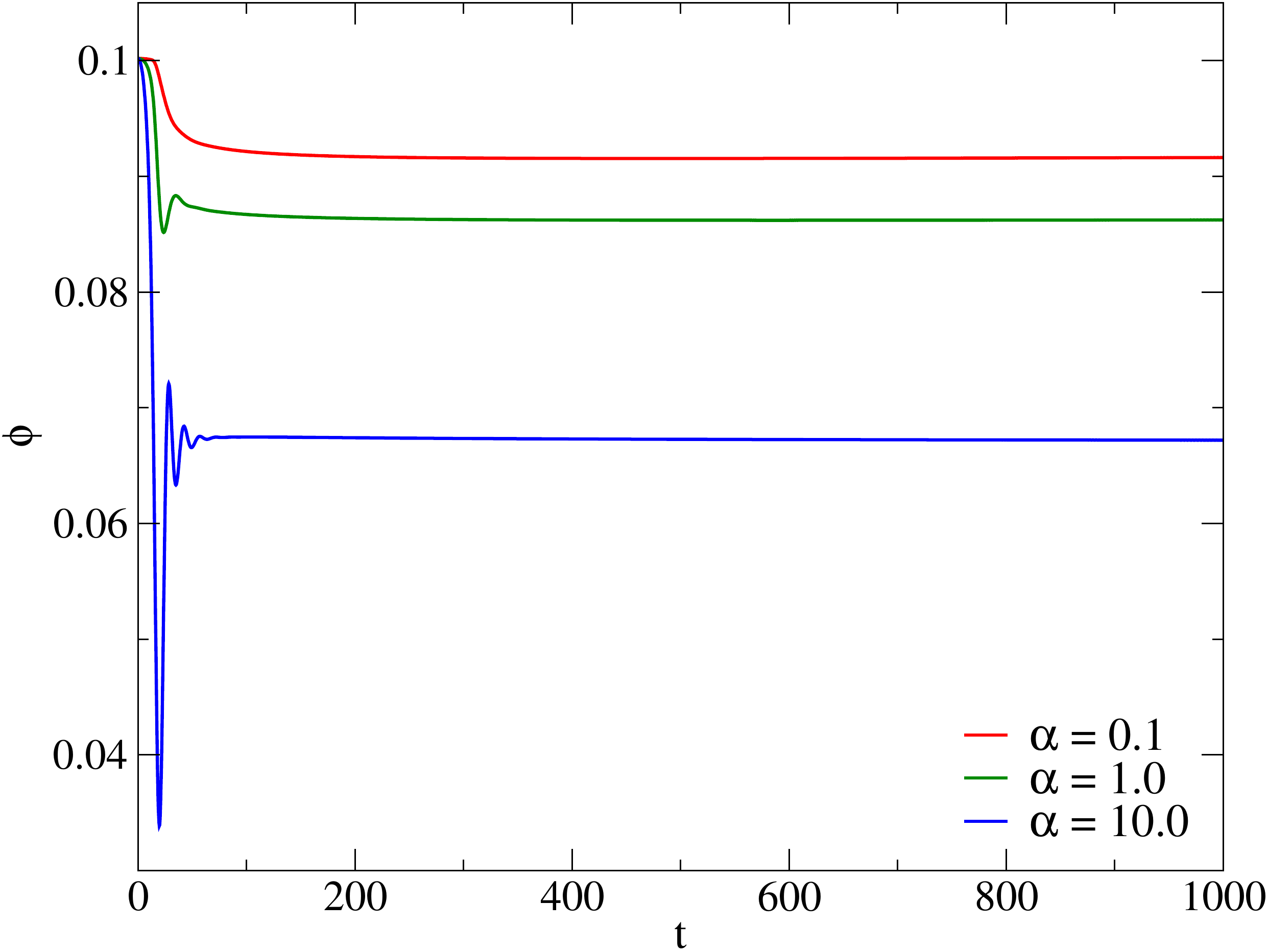}
\\
\caption{(Left panel) Radial profile of the scalar field $\phi(r)$ at $t=1000$ for the initial data consisting of a GMGHS BH with $q=1.1$, in three models with different $\alpha$. There is an excellent matching between the profile obtained from the numerical evolution and the profile of a AAM BH with a slightly larger $q$ and smaller $Q_s$. (Right panel) Time evolution of the amplitude of the scalar field extracted at $r_0 =  11.09$.}
\label{fig4b}
\end{figure}

Let us now describe the results obtained within this setup starting with the first sort of initial data described above. 
In Fig.~\ref{fig4} (left panel)
we exhibit the radial profile of the scalar field at late times ($t=600$),  
and we compare it with the corresponding analytic profile of the AAM solution described in Section~\ref{sec2} with the corresponding value of $q$ and $r_H$. 
 In the  right panels we show the time evolution of the BH charge $Q_{e}$ and the amplitude of the scalar field extracted at radius $r_0 =  11.09$. 
These quantities clearly stabilise reaching an equilibrium configuration, which matches the AAM solution.  At the end of the evolution, the final charge is $Q_{e}=0.196$.

To assess the dynamical stability of the AAM without imposing spherical symmetry, we carried out numerical simulations in a 3D cartesian grid using the \textsc{Einstein Toolkit}~\cite{EinsteinToolkit,Loffler}. To perform the evolutions we have used a numerical grid with 11 refinement levels with 
\begin{eqnarray} 
\lbrace &&(192, 96, 48, 24, 12, 6, 3, 1.5, 0.75, 0.375, 0.1875)
\\
\nonumber
 &&
(6.4, 3.2, 1.6, 0.8, 0.4, 0.2, 0.1, 0.05, 0.025, 0.0125, 0.00625)~~\rbrace \ , 
\end{eqnarray} 
 where the first set of numbers indicates the spatial domain of each level and the second set indicates the resolution. The evolution is identical to the spherically symmetric case. In Fig.~\ref{fig4n} we plot snapshots of the evolution of the scalar field. Similar diagnostics as for the 1+1 evolutions attest the convergence of the initial data towards an AAM BH.

The second sort of initial data is dealt with in a very similar way. We now start with a GMGHS solution with $q=1.1$. We consider three models with $\alpha=0.1, 1, 10$. The evolution changes the value of the scalar field and increases slightly the value of the charge to mass ratio of the BH. We obtain, respectively, $q=1.1001, 1.1001, 1.101$. The radial profile of the scalar field can be matched with a AAM BH with, respectively, $Q_s=1.10, 1.03, 0.8$. The profiles of the scalar field at $t=1000$ and the evolution (left panel) of the evolution of the scalar field at some extraction radius (right panel) can be seen in Fig.~\ref{fig4b}.

These evolutions agree with our expectations. Starting from RN, but forcing the system to evolve to an AAM solutions, due to the dilaton coupling, the BH grows a scalar charge. Starting from GMGHS, but forcing the system to evolve to an AAM BH due to the non-trivial potential, the BH loses some scalar charge. The first/second case establishes the dynamical formation of an undercharged/overcharged AAM BH.

\section{Thermodynamical stability}
\label{sec5}
As pointed out in~\cite{Astefanesei:2019mds},
a unique property (within asymptotically flat spacetime BHs) of AAM BHs is that
they possess a subset which is locally  thermodymically stable. We shall now examine precisely when this occurs in the domain of existence. 

 Thermodynamical stability can be {\it local} or {\it global}. 
Moreover, different thermodynamic ensembles represent physically different situations and may not lead, in general, to the same conclusions regarding the thermodynamical stability. Mathematically, (local) thermodynamical
stability is equated with the sub-additivity of the entropy function. 
In the canonical ensemble, this is equivalent to the positivity of the 
specific heat at constant electric charge
\begin{equation} 
C_Q  =T_H \left( \frac{\partial S}{\partial T_H} \right)\bigg|_{Q_e} \geqslant 0 \ .
\end{equation}
 In the grand canonical ensemble, one requires instead the positivity of the specific heat at constant electric potential, 
and the positivity of the isothermal permittivity 
\begin{equation} 
C_\Phi =T_H \left(\frac{\partial S}{\partial T_H} \right)\bigg|_{\Phi} \geqslant 0 \ , \qquad
{\rm and}~~~~
\epsilon_T=\left(\frac{\partial Q_e}{\partial \Phi} \right)\bigg|_{T_H} \geqslant 0 \ .
\end{equation}
In fact, if $C_Q$ and $\epsilon_T$ are positive,
the identity
\begin{equation} 
C_\Phi =C_Q +T_H \epsilon_T \alpha_Q^2  \ , \qquad {\rm where } \ \ \ \alpha_Q\equiv  \left(\frac{\partial \Phi}{\partial T_H}\right)\bigg|_{Q_e} \ ,
\end{equation}
implies $C_\Phi>0$. Thus, local thermodynamical stability follows from $C_Q\geqslant 0$ and $\epsilon_T\geqslant 0$. 

For the Schwarzschild BH there are no electric variables and the specific heat is negative
\begin{eqnarray}
\label{Scw5}
C=-8\pi M^2<0 \ ,
\end{eqnarray}
which means it is locally thermodynamically unstable. For the RN BH, parameterised in terms of $r_H, Q$, BH solutions exist for $Q_e\leqslant r_H$.
The specific heats are
\begin{eqnarray} 
C_Q=\frac{2\pi r_H^2(r_H^2-Q_e^2)}{3Q_e^2-r_H^2} \ , \qquad C_\Phi=-2\pi r_H^2<0 \ ,
\end{eqnarray}
whereas the electric permittivity is
\begin{eqnarray} 
\epsilon_T=-\frac{r_H(3Q_e^2-r_H^2)}{r_H^2-Q_e^2} \ .
\end{eqnarray}
Thus, RN BHs exhibit two phases, depending on the sign of $r_H^2-Q_e^2$, which vanishes for $q=\sqrt{3}/2$. For $q<\sqrt{3}/2$, $C_Q<0$ and $\epsilon_T>0$. This is the Schwarzschild-like phase. For $q>\sqrt{3}/2$, $C_Q>0$ and $\epsilon_T<0$. This is the near  extremal RN-like phase.  
RN BHs are stable in the canonical ensemble in the near extremal RN-like phase. But they are always unstable in the grand canonical ensemble. Notice, however, that $C_Q$ is vanishing at extremality, wherein $\epsilon_T$ is diverging. Thus, the RN family seems to be approaching a new phase at extremality, wherein BHs cease to be possible.

For the GMGHS family with $\gamma=1$, parameterised by $M,r_H$, the response functions take the simple form
\begin{eqnarray} 
C_Q=-8\pi M^2<0 \ , \qquad C_\Phi=-2\pi r_H^2<0\ , \qquad \epsilon_T=2M>0 \ .
\end{eqnarray}
The negativity of the specific heats implies local thermodynamical instability.

\begin{figure}[t]
\centering
\includegraphics[width=0.49\linewidth,angle=0]{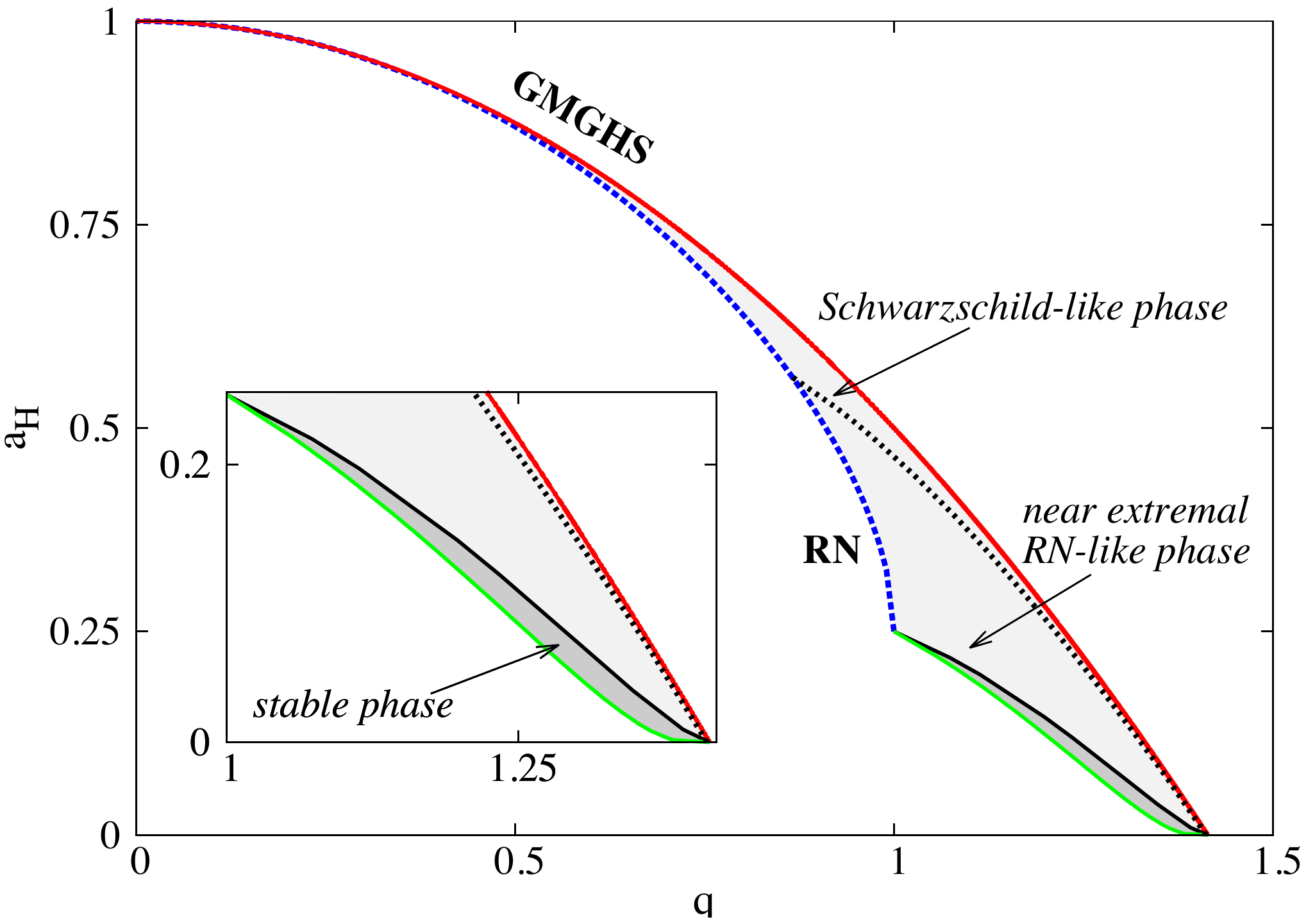}  
\includegraphics[width=0.49\linewidth,angle=0]{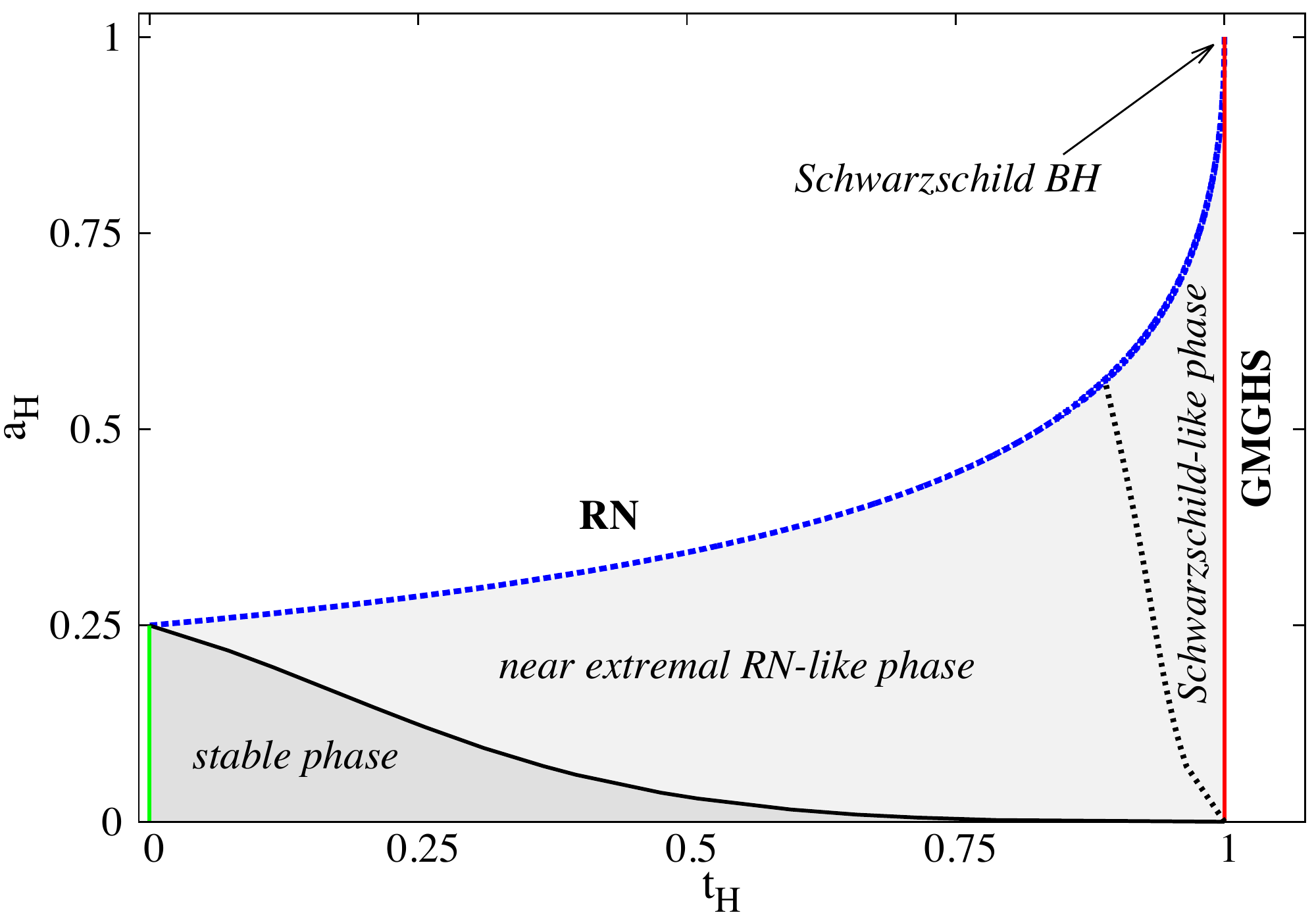} 
\caption{The three thermodynamical phases of AAM BHs plotted in the same representations of the domain of existence shown in Fig.~2.
}
\label{Fig6}
\end{figure}

In the AAM case, the expression of 
$C_Q$, $C_\Phi$, and $\epsilon_T$ are long and not enlightening \cite{Astefanesei:2019mds}, thus we do not include them here. All the response functions can, nonetheless, be presented as functions of $r_H$ and $Q_e$.
A study of these quantities show the existence of a region in the domain of existence
where the thermodynamic stability is satisfied in both 
canonical and grand-canonical ensembles.
This region is bounded by the set of extremal solutions 
$T_H=0$
and a set of $critical$ configurations  where $C_\Phi$
and $\epsilon_T$
diverge (and change sign afterwards)
while $C_Q$
remains positive and finite.
The   critical  configurations are found numerically,
the relations
\begin{eqnarray} 
\nonumber
&&
a_H =  \left(\frac{ q - \sqrt{2} }{ 1 - \sqrt{2} } \right)
	\bigg(\frac{1}{4}
	+ (0.220\pm 0.001)(q - 1) + 
	(0.12\pm 0.05)(q - 1)^2 
	\\
	\nonumber
	&&
					{~~~~~~~~~~~~~~~~~~~~~~~~}
	- (5.5\pm 0.5)(q - 1)^3 + (23\pm 2)(q - 1)^4 
	- (41\pm 2)(q - 1)^5
				\bigg) \ ,
				\\
				\nonumber
				&&
a_H=(1 - t_H)^3 
\bigg(
\frac{1}{4} + (
    0.3807\pm 0.0007)t_H 
		- (0.621\pm 0.009)t_H^2 
		- (1.34\pm 0.04) t_H^3 
		\\
	\nonumber
	&&
	{~~~~~~~~~~~~~~~~~~~~~~}
	  + (2.12\pm 0.05)t_H^4
			\bigg)~.{~~}
\end{eqnarray}
providing a good fit for the corresponding curves in Fig.~\ref{Fig6}. 
Let us also remark that all thermodynamically stable AAM BHs have
$1<q<\sqrt{2}$,
while
 $\sqrt{2}/2<\Phi<1$.

\begin{figure}[t]
\centering
\includegraphics[width=0.49\linewidth,angle=0]{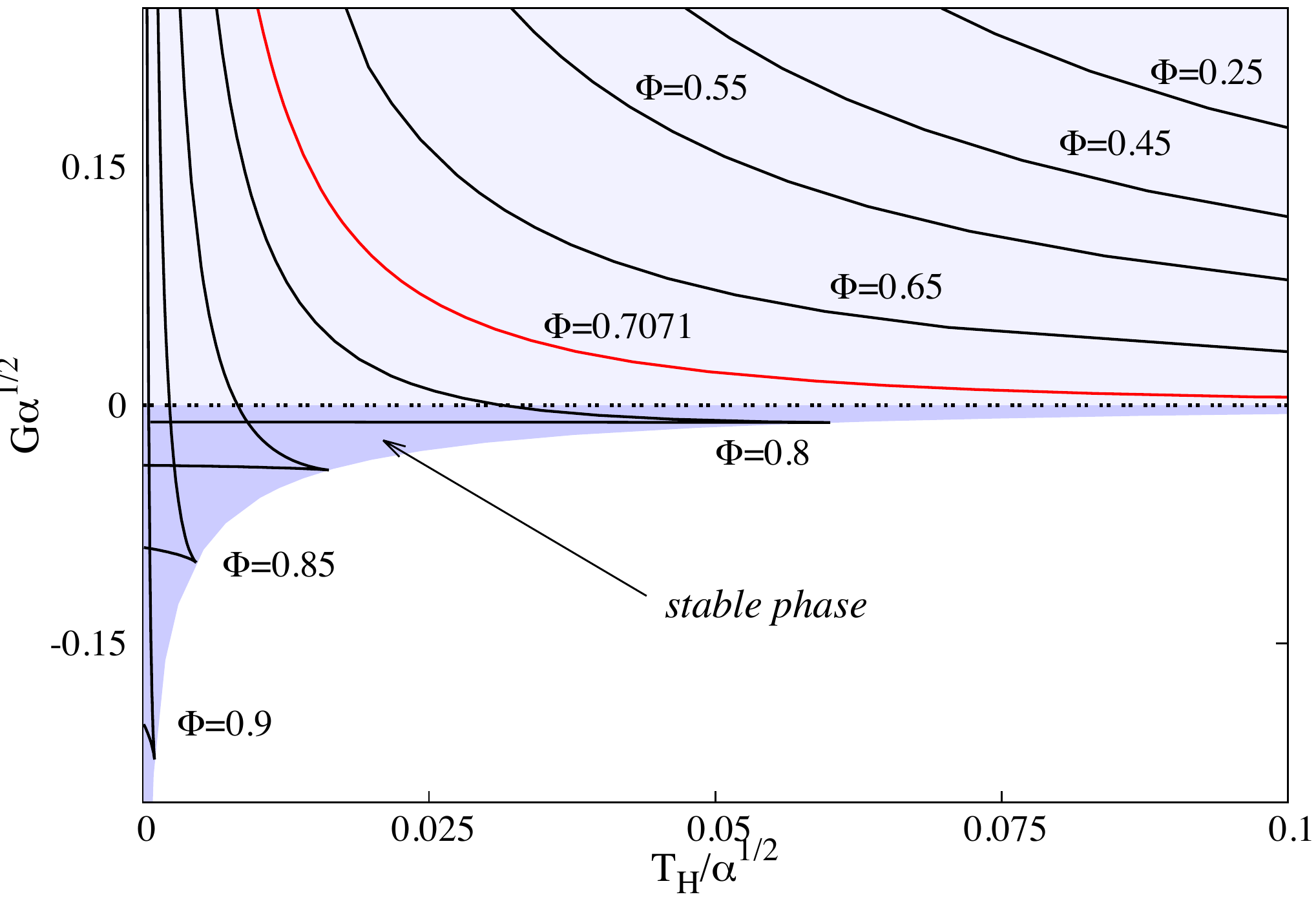} 
\includegraphics[width=0.49\linewidth,angle=0]{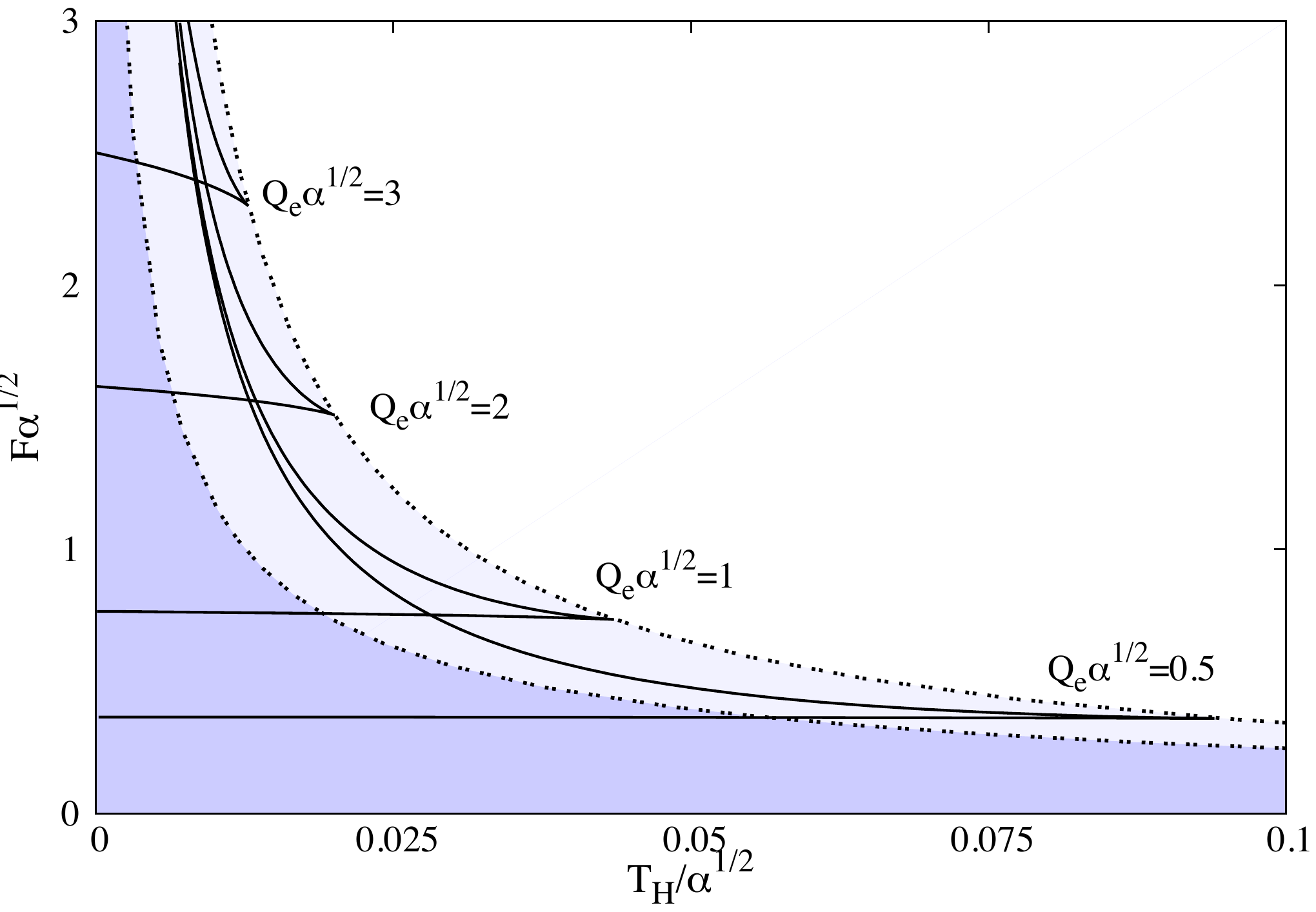}  
\caption{ (Left panel)  The Gibbs  free energy $vs.$ temperature for fixed chemical potential ensemble. 
(Right panel)  The Helmholtz free energy $vs.$ temperature for the fixed charge ensemble.
In both plots, the BHs exist in the light blue region only,
 the shaded blue region corresponding  to stable phase solutions.
}
\label{Fig7}
\end{figure}

Moreover, the set of  {\it locally} thermodynamically stable solutions
are also  {\it globally} stable.
That is, in a grand canonical ensemble, 
($i.e.$ for the same $T_H, \Phi$)
they minimise the Gibbs free energy
\begin{eqnarray}
G=M-T_H S-\Phi Q_e~.
\end{eqnarray}
The generic picture is summarised in Fig.~\ref{Fig7} (left panel).
For any value of  $1/\sqrt{2}<\Phi<1$,
 the $G(T_H)$ curve 
consists in two parts.
The branch minimising the free energy $G$  
  starts at $T_H=0$ and ends at some maximal $T_H(\Phi)$,
consisting in configurations which are {\it locally} stable.
The situation changes for 
$\Phi<1/\sqrt{2}$,
in which case, similar to the RN or GMGHS cases,
one single branch of locally unstable solutions is found\footnote{One finds
 \begin{eqnarray}
G=\frac{(1-\Phi^2)^2}{16\pi T_H}~~{\rm for~RN~BHs},~~{\rm and}~~
G=\frac{1-2\Phi^2 }{16\pi T_H}~~{\rm for~GMGHS~BHs},
\end{eqnarray}
while the corresponding expression for AAM BHs cannot be explicitly written in terms of $\Phi,T_H$.
},
with $G>0$.

When considering instead a canonical ensemble,
one finds 
the existence of two branches,
for any (fixed) value of the electric charge
$Q_e$.
The solutions minimising the Helmholtz
free energy
  \begin{eqnarray}
F=M-T_H S~,
\end{eqnarray}
are located on the lower branch, 
which starts with the  $T_H=0$ extremal BHs.
These configurations have also a positive  specific heat, $C_Q>0$,
while a part of them are also stable in a grand canonical ensemble, $C_\Phi>0$.

To summarise, we conclude that a set of AAM solutions 
which are overcharged, $q>1$, and
with a large enough chemical potential, $\Phi>1/\sqrt{2}$,
are 
thermodynamically stable, both
 {\it locally} and {\it globally}.

\section{Conclusions}
\label{sec6}

In this paper we have discussed the set of BH solutions found in~\cite{Anabalon:2013qua} within Einstein-Maxwell-dilaton theory with a certain dilaton potential. Such  AAM BHs can be thought of as a family of solutions that interpolates between the standard RN electrovacuum BHs and the GMGHS solutions of low energy heterotic string theory in four dimensions, retaining some of the features of both these limits. In particular, these BHs have a regular extremal limit and no electric charge outside the horizon, analogously to the RN BH; on the other hand, they can be overcharged, i.e. to have a charge to mass ratio exceeding unity, as GMGHS. The combination of these properties allows in particular the exceptional feature in BH physics of exhibiting thermodynamical stability in both the canonical and grand canonical ensemble. In this sense, the overcharged AAM BHs can be faced as an extension of the RN family beyond extremality.

Although there is a subset of AAM BHs that have both dynamical and thermodynamical stability, they are still afflicted by the decay induced by quantum effects, that is, Hawking radiation, except for the extremal solutions. The extremal AAM BHs are then stable also against Hawking evaporation. One cannot exclude, however, if these solutions are not supersymmetric, that non-perturbative effects may destabilise them. 
 
We would like also to point that similar results in the thermodynamical analysis have been found for a second model discussed in~\cite{Astefanesei:2019mds}, still described by (\ref{action}), but with $\gamma=\sqrt{3}$ and a different $V(\phi)$. Clearly, some of the analysis herein could be repeated for that model. 

Finally, despite the ingredients we have identified, we cannot pinpoint exactly the mechanism behind the existence of these dynamically and thermodynamically stable BHs. In particular the properties of the potential that permit them to exist. It is well known that $AdS$ BHs can become thermodynamically stable. It is then tempting to think the dilaton potential is inducing $AdS$-like features. There is, however, an important difference between these two cases. In the $AdS$ case, large BHs are thermodynamically stable; in the case analysed herein, the stable BHs are the smallest ones.
In this respect it is worth remarking that the potential~(\ref{V}) is decaying towards the spatial infinity in the AAM BH solutions. Thus, even if induces a box-like effect, such effect may be more effective for small BHs.

\section*{Acknowledgements}
The work of DA is supported by the Fondecyt Regular Grants 1161418 and 1171466. The work of E.R. is supported by the Fundacao para a Ci\^encia e a Tecnologia (FCT) project UID/MAT/04106/2019 (CIDMA) and by national funds (OE), through FCT, I.P., in the scope of the framework contract foreseen in the numbers 4, 5 and 6
of the article 23, of the Decree-Law 57/2016, of August 29,
changed by Law 57/2017, of July 19. 
J.L.B.S. would like to acknowledge support from DFG Research Training Group 1620  \textit{Models of Gravity} and
the DFG Project No. BL 1553. 
We acknowledge support  from the project PTDC/FIS-OUT/28407/2017.  
 This work has further been supported by  the  European  Union's  Horizon  2020  research  and  innovation  (RISE) programmes H2020-MSCA-RISE-2015
Grant No.~StronGrHEP-690904 and H2020-MSCA-RISE-2017 Grant No.~FunFiCO-777740. 
The authors would like to acknowledge networking support by the
COST Action CA16104. 
%


	

\begin{thebibliography}{99}
 
\bibitem{Israel:1967wq}
  W.~Israel,
  Phys.\ Rev.\  {\bf 164} (1967) 1776.
\bibitem{Regge:1957td}
  T.~Regge and J.~A.~Wheeler,
  Phys.\ Rev.\  {\bf 108} (1957) 1063.
\bibitem{Zerilli:1970se}
  F.~J.~Zerilli,
  Phys.\ Rev.\ Lett.\  {\bf 24} (1970) 737.
\bibitem{Hawking:1974sw}
  S.~W.~Hawking,
  Commun.\ Math.\ Phys.\  {\bf 43} (1975) 199
   Erratum: [Commun.\ Math.\ Phys.\  {\bf 46} (1976) 206].
\bibitem{Hawking:1976de}
  S.~W.~Hawking,
  Phys.\ Rev.\ D {\bf 13} (1976) 191.
\bibitem{Israel:1967za}
  W.~Israel,
  Commun.\ Math.\ Phys.\  {\bf 8} (1968) 245.
\bibitem{Moncrief:1974gw}
  V.~Moncrief,
  Phys.\ Rev.\ D {\bf 9} (1974) 2707.
\bibitem{Moncrief:1974ng}
  V.~Moncrief,
  Phys.\ Rev.\ D {\bf 10} (1974) 1057.
  \bibitem{Davies:1977}
  P.~C.~W.~Davies,
  Proc.\ R.\ Soc. \ Lond. A {\bf 353} (1977) 499.
\bibitem{Gibbons:1982ih}
  G.~W.~Gibbons,
  Nucl.\ Phys.\ B {\bf 207} (1982) 337.
\bibitem{Gibbons:1987ps}
  G.~W.~Gibbons and K.~i.~Maeda,
  Nucl.\ Phys.\ B {\bf 298} (1988) 741.
\bibitem{Garfinkle:1990qj} 
D.~Garfinkle, G.~T.~Horowitz and A.~Strominger,
Phys.\ Rev.\ D {\bf 43}, 3140 (1991)
Erratum: [Phys.\ Rev.\ D {\bf 45}, 3888 (1992)].
  
  
\bibitem{Anabalon:2017yhv}
  A.~Anabalon, D.~Astefanesei, A.~Gallerati and M.~Trigiante,
  JHEP {\bf 1804} (2018) 058
  [arXiv:1712.06971 [hep-th]].
\bibitem{Anabalon:2013qua}
  A.~Anabalon, D.~Astefanesei and R.~Mann,
  JHEP {\bf 1310} (2013) 184
  [arXiv:1308.1693 [hep-th]].
\bibitem{Astefanesei:2019mds}
  D.~Astefanesei, D.~Choque, F.~Gómez and R.~Rojas,
  JHEP {\bf 1903} (2019) 205
  [arXiv:1901.01269 [hep-th]].
\bibitem{Blazquez-Salcedo:2019nwd}
  J.~L.~Blazquez-Salcedo, S.~Kahlen and J.~Kunz,
  arXiv:1911.01943 [gr-qc].
\bibitem{Jansen:2019wag}
  A.~Jansen, A.~Rostworowski and M.~Rutkowski,
  arXiv:1909.04049 [hep-th].
\bibitem{Herdeiro:2018wub}
  C.~A.~R.~Herdeiro, E.~Radu, N.~Sanchis-Gual and J.~A.~Font,
  Phys.\ Rev.\ Lett.\  {\bf 121} (2018) no.10,  101102
  [arXiv:1806.05190 [gr-qc]].
\bibitem{Fernandes:2019rez}
  P.~G.~S.~Fernandes, C.~A.~R.~Herdeiro, A.~M.~Pombo, E.~Radu and N.~Sanchis-Gual,
  Class.\ Quant.\ Grav.\  {\bf 36} (2019) no.13,  134002
  [arXiv:1902.05079 [gr-qc]].
\bibitem{Fernandes:2019kmh}
  P.~G.~S.~Fernandes, C.~A.~R.~Herdeiro, A.~M.~Pombo, E.~Radu and N.~Sanchis-Gual,
  Phys.\ Rev.\ D {\bf 100} (2019) no.8,  084045
  [arXiv:1908.00037 [gr-qc]].

\bibitem{Misner:1964je}
  C.~W.~Misner and D.~H.~Sharp,
  Phys.\ Rev.\  {\bf 136} (1964) B571.
 
\bibitem{Astefanesei:2018vga} 
  D.~Astefanesei, R.~Ballesteros, D.~Choque and R.~Rojas,
  Phys.\ Lett.\ B {\bf 782}, 47 (2018)
  doi:10.1016/j.physletb.2018.05.005
  [arXiv:1803.11317 [hep-th]].
 
 
\bibitem{Herdeiro:2015waa}
  C.~A.~R.~Herdeiro and E.~Radu,
  Int.\ J.\ Mod.\ Phys.\ D {\bf 24} (2015) no.09,  1542014
  [arXiv:1504.08209 [gr-qc]].
 
\bibitem{Goldstein:2005hq} 
  K.~Goldstein, N.~Iizuka, R.~P.~Jena and S.~P.~Trivedi,
  Phys.\ Rev.\ D {\bf 72}, 124021 (2005)
  [hep-th/0507096].
 

\bibitem{Sen:2005wa} 
  A.~Sen,
  JHEP {\bf 0509}, 038 (2005)
  [hep-th/0506177].
  
\bibitem{Astefanesei:2006dd} 
  D.~Astefanesei, K.~Goldstein, R.~P.~Jena, A.~Sen and S.~P.~Trivedi,
  JHEP {\bf 0610}, 058 (2006)
  [hep-th/0606244].
 
\bibitem{Sen:2007qy} 
  A.~Sen,
  Gen.\ Rel.\ Grav.\  {\bf 40}, 2249 (2008)
  [arXiv:0708.1270 [hep-th]].
 
\bibitem{Anabalon:2013sra} 
  A.~Anabalón and D.~Astefanesei,
  Phys.\ Lett.\ B {\bf 727}, 568 (2013)
  [arXiv:1309.5863 [hep-th]].
\bibitem{Robinson:1959ev}
  I.~Robinson,
  Bull.\ Acad.\ Pol.\ Sci.\ Ser.\ Sci.\ Math.\ Astron.\ Phys.\  {\bf 7} (1959) 351.
\bibitem{Bertotti:1959pf}
  B.~Bertotti,
 Phys.\ Rev.\  {\bf 116} (1959) 1331.

\bibitem{Blazquez-Salcedo:2018jnn} 
J.~L.~Bl\'azquez-Salcedo, D.~D.~Doneva, J.~Kunz and S.~S.~Yazadjiev,
Phys.\ Rev.\ D {\bf 98}, no. 8, 084011 (2018)
[arXiv:1805.05755 [gr-qc]].
 
 
\bibitem{Kokkotas:1999bd} 
K.~D.~Kokkotas and B.~G.~Schmidt,
Living Rev.\ Rel.\  {\bf 2}, 2 (1999)
[gr-qc/9909058].
\bibitem{Nollert:1999ji} 
H.~P.~Nollert,
Class.\ Quant.\ Grav.\  {\bf 16}, R159 (1999).
\bibitem{Berti:2009kk} 
E.~Berti, V.~Cardoso and A.~O.~Starinets,
Class.\ Quant.\ Grav.\  {\bf 26}, 163001 (2009)
[arXiv:0905.2975 [gr-qc]].
\bibitem{Konoplya:2011qq} 
R.~A.~Konoplya and A.~Zhidenko,
Rev.\ Mod.\ Phys.\  {\bf 83}, 793 (2011)
[arXiv:1102.4014 [gr-qc]].
 
\bibitem{Ferrari:2000ep} 
V.~Ferrari, M.~Pauri and F.~Piazza,
Phys.\ Rev.\ D {\bf 63}, 064009 (2001)
[gr-qc/0005125].

\bibitem{Blazquez-Salcedo:2018pxo} 
J.~L.~Bl\'azquez-Salcedo, Z.~Altaha Motahar, D.~D.~Doneva, F.~S.~Khoo, J.~Kunz, S.~Mojica, K.~V.~Staykov and S.~S.~Yazadjiev,
Eur.\ Phys.\ J.\ Plus {\bf 134}, no. 1, 46 (2019)
[arXiv:1810.09432 [gr-qc]].

 
\bibitem{Leaver:1990zz} 
E.~W.~Leaver,
Phys.\ Rev.\ D {\bf 41}, 2986 (1990).
	

\bibitem{Blazquez-Salcedo:2016enn} 
J.~L.~Bl\'azquez-Salcedo, C.~F.~B.~Macedo, V.~Cardoso, V.~Ferrari, L.~Gualtieri, F.~S.~Khoo, J.~Kunz and P.~Pani,
Phys.\ Rev.\ D {\bf 94}, no. 10, 104024 (2016)
[arXiv:1609.01286 [gr-qc]].
\bibitem{Blazquez-Salcedo:2017txk} 
J.~L.~Bl\'azquez-Salcedo, F.~S.~Khoo and J.~Kunz,
Phys.\ Rev.\ D {\bf 96}, no. 6, 064008 (2017)
[arXiv:1706.03262 [gr-qc]].

\bibitem{Sanchis-Gual:2015lje}
  N.~Sanchis-Gual, J.~C.~Degollado, P.~J.~Montero, J.~A.~Font and C.~Herdeiro,
  Phys.\ Rev.\ Lett.\  {\bf 116} (2016) no.14,  141101
  [arXiv:1512.05358 [gr-qc]].
\bibitem{Sanchis-Gual:2016tcm}
  N.~Sanchis-Gual, J.~C.~Degollado, C.~Herdeiro, J.~A.~Font and P.~J.~Montero,
  Phys.\ Rev.\ D {\bf 94} (2016) no.4,  044061
  [arXiv:1607.06304 [gr-qc]].
\bibitem{Hirschmann:2017psw}
  E.~W.~Hirschmann, L.~Lehner, S.~L.~Liebling and C.~Palenzuela,
  Phys.\ Rev.\ D {\bf 97} (2018) no.6,  064032
  [arXiv:1706.09875 [gr-qc]].
\bibitem{CorderoCarrion:2012ic}
  I.~Cordero-Carrion and P.~Cerda-Duran,
  arXiv:1211.5930 [math-ph].
\bibitem{Cordero2} Cordero-Carrión, I., \& Cerdá-Durán, P. (2014). Partially Implicit Runge-Kutta Methods for Wave-Like Equations. In Advances in Differential Equations and Applications (pp. 267-278). Springer, Cham.
\bibitem{EinsteinToolkit} EinsteinToolkit, Einstein Toolkit: Open software for relativistic
astrophysics, http://einsteintoolkit.org/.
\bibitem{Loffler} L\"offler, F. et al. (2012). The Einstein Toolkit: a community computational infrastructure for relativistic astrophysics. Classical and Quantum Gravity, 29 (11), 115001.
 



  

	
\end{thebibliography}
\end{document}